\documentclass[a4paper,fleqn,usenatbib]{mnras}

\usepackage[T1]{fontenc}
\usepackage{ae,aecompl}
\usepackage{color}

\usepackage{graphicx}	% Including figure files
\usepackage{amsmath}	% Advanced maths commands
\usepackage{amssymb}	% Extra maths symbols
\usepackage{txfonts}

\newcommand{\ie}{{i.e. \/}}
\newcommand{\mtl}{\ensuremath{M_*/L \ }}
\newcommand{\Msun}{\ensuremath{{\rm M}_{\odot}}}
\title[A solution to the missing mass problem]{Spatially-unresolved SED fitting can underestimate galaxy masses: a solution to the missing mass problem}

\author[R. Sorba and M. Sawicki]{
Robert Sorba,$^{1,2}$\thanks{E-mail: rsorba@mta.ca, marcin.sawicki@smu.ca}
Marcin Sawicki$^{1}$\thanks{Canada Research Chair}
\\
$^{1}$Department of Astronomy and Physics, and Institute for Computational Astrophysics, Saint Mary's University, 923 Robie Street, Halifax, Nova Scotia, B3H 3C3, Canada\\
$^{2}$Physics Department, Mount Allison University, 62 York Street, Sackville, New Brunswick, E4L 1E2, Canada\\
}

\date{Accepted XXX. Received YYY; in original form ZZZ}

\pubyear{2017}
\begin{document}
\label{firstpage}
\pagerange{\pageref{firstpage}--\pageref{lastpage}}
\maketitle

\begin{abstract}
{\color{black} We perform spatially-resolved, pixel-by-pixel SED fitting on galaxies up to $z\sim2.5$ in the Hubble Extreme Deep Field (XDF). Comparing stellar mass estimates from spatially resolved and spatially unresolved photometry 
we find that unresolved masses can be systematically underestimated by factors of up to 5. The ratio of the unresolved to resolved mass measurement depends on the galaxy's specific star formation rate (sSFR): at low sSFRs the bias is small, but above sSFR$\ \sim 10^{-9.5}$ yr$^{-1}$ the discrepancy increases rapidly such that galaxies with sSFRs$\ \sim 10^{-8}$ yr$^{-1}$ have unresolved mass estimates of only one half to one fifth of the resolved value. This result indicates that stellar masses estimated from spatially-unresolved datasets need to be systematically corrected, in some cases by large amounts, and we provide an analytic prescription for applying this correction. We show that correcting stellar mass measurements for this bias changes the normalization and slope of the star-forming main sequence and reduces its intrinsic width; most dramatically, correcting for the mass bias increases the stellar mass density of the Universe at high redshift and can resolve the long-standing discrepancy between the directly-measured cosmic star formation rate density at $z\ga1$ and that inferred from stellar mass densities ("the missing mass problem").}
\end{abstract}


%%%%%%%%%%%%%%%%%%%%%%%%%%%%%%%%%%%%%%%%%%%%%%%%%%

%%%%%%%%%%%%%%%%% BODY OF PAPER %%%%%%%%%%%%%%%%%%

\section{Introduction}
\label{pxp2:intro}

Spatially resolved (\ie pixel-by-pixel) broadband SED fitting provides an exciting tool to study the spatial distribution of stellar matter in a galaxy, particularly for high redshift galaxies where obtaining 2-D spectra is cost prohibitive. Resolving features in a galaxy's spatial distribution allows us to uncover details that are obscured and blotted out by integrated photometry. The technique of pixel-by-pixel SED fitting was first developed by \citet{Abraham1999}, who matched spectral synthesis models to resolved multicolor data of $z \sim 1$ galaxies in the Hubble Deep Field in order to study ages and evolutionary histories of stellar populations in those galaxies. An analysis of how pixel-by-pixel fitting affected mass estimates was performed by \citet{Zibetti2009} for {\color{black} 9} nearby ($D < 30$ Mpc) SINGS galaxies. They {\color{black} used median stellar mass-to-light (\mtl) ratios (derived from a large SPS library) to create stellar mass maps of the galaxies} and found that the unresolved mass estimate (i.e. from integrating all the light of the galaxy) could be underestimated compared to the resolved mass by up to 40\% (0.22 dex). They postulated that the presence of large dust lanes contributed to the mass discrepancy. \citet{Sorba2015} corroborated this by using pixel-by-pixel analysis on a larger set of 67 nearby galaxies in the Sloan Digital Sky Survey \citep[SDSS;][]{Eisenstein2011}. They found that the presence of strong dust lanes could cause unresolved mass estimates to be underestimated by 45\% (0.25 dex) for an individual galaxy, but only in extreme cases, and that dust did not affect mass estimates on average. However, by being able to disentangle older stellar populations from younger ones, \citet{Sorba2015} were able to find an increasing bias in a galaxy's mass estimation with specific star-formation rate (sSFR). They found the effect was small (unresolved masses were underestimated by 13\% or 0.06 dex on average), but strongly correlated with sSFR leading to a maximum bias of 25\% (0.12 dex) at sSFRs of around $10^{-8} yr^{-1}$, and they concluded this bias was due to outshining. Essentially, young massive stars have a much smaller \mtl than older stars. They are orders of magnitude brighter than solar type stars. But they are also short-lived, meaning they are only present in currently star-forming regions. When all the light from a galaxy is integrated together, however, the light from young stars dominates a galaxy's SED at optical wavelengths. Thus model SED fits to broadband photometry preferentially fit the \mtl of the younger stellar population, often missing mass from older components \citep[for further discussion on outshining and its effects, see][]{Sawicki1998, Papovich2001, Maraston2010, Pforr2012}.

Spatially resolved stellar property maps have also been created for high redshift ($0.5 < z < 2.5$) star-forming galaxies in the Cosmic Assembly Near-infrared Deep Extragalactic Legacy Survey \citep[CANDELS;][]{Grogin2011, Koekemoer2011} by \citet{Wuyts2012}. They used the maps to examine star-forming clumps and variations in rest-frame color, stellar surface mass density, age, and extinction as a function of radius, finding results consistent with inside-out disk-growth. They also compared unresolved to pixel-by-pixel mass estimates, but found no discrepancy. However, their binning of low signal-to-noise (S/N) pixels may have impacted their effective spatial resolution, leaving an open question as to how outshining may change at higher redshifts. In this work we set out to examine the influence of outshining as a function of cosmic time. We do so following the pixel-by-pixel method laid out in \citet{Sorba2015}. {\color{black} It is important to note that this work is a comparative study between the results given by pixel-by-pixel SED-fitting versus the standard integrated-light SED fitting approach. When we discuss any bias in mass estimation, this is only in relation to how the two methods compare. We do not presume to know, and for our purposes do not need to know, the true stellar mass of the galaxies studied.} 

Throughout this work we assume a WMAP7 \citep{Komatsu2011} flat $\Lambda$CDM cosmology ($H_0 = 70.4$ km s$^{-1}$ Mpc$^{-1}$, $\Omega_M = 0.272$) and use the AB magnitude system.

\section{Method}
\label{pxp2:method}

\subsection{Data}
\label{pxp2:data}

To examine pixel-by-pixel SED fitting at higher redshifts we used the Hubble eXtreme Deep Field \citep[XDF;][]{Illingworth2013} publicly available dataset, which combines data from a decade of observations made with the Hubble Space Telescope (HST) creating optical/NIR images unsurpassed in depth. The extreme depth of this dataset is particularly important for a pixel-by-pixel analysis, increasing the number of pixels with acceptable signal-to-noise ($S/N$) and enabling us to probe fainter regions of galaxies. The XDF dataset contains mosaic images covering approximately four square arcminutes in nine passbands (ACS/WFC $F435W, F606W, F775W, F814W$, and $F850LP$; WFC3/IR $F105W, F125W, F140W$, and $F160W$) that have all had their background flux removed, been astrometrically aligned, and drizzled onto the same 60 mas pixel scale.

It is important when doing pixel-by-pixel photometry that all images have the same point spread function (PSF). To measure the PSF of each bandpass, we started by visually inspecting stars from the catalog of stars in the Hubble Ultra-Deep Field \citep{Pirzkal2005} and selecting those that weren't contaminated by noise or nearby galaxies. These stars were re-centered to their intensity-weighted sub-pixel centers, normalized, and median-stacked to form the PSF. The WFC3 PSFs were augmented at larger radii using model PSFs created using the TinyTim software package \citep{Krist1995, Krist2011} to account for the relatively large amount of flux contained in the broad wings of the NIR PSFs \citep{VanderWel2012, Rafelski2015}. The broadest PSF belonged to the $F160W$ image, which had a PSF full-width-at-half-maximum (FWHM) of approximately 0.20 arcseconds. For each bandpass other than $F160W$, we convolved the images to match the broadest PSF by creating an $11\times11$ convolution kernel using the method described in \citet{Alard1998}, employing a delta function at each pixel in the kernel as the set of basis functions \citep{Becker2012}. The size of the kernel provided a good compromise, limiting the tendency of our choice of basis function to over-fit the noise in the wings of the PSF, but was still large enough to capture all the features of the transformation kernel. 

After convolving the images to the same PSF, we created a multi-band detection image by summing each image weighted by the inverse variance on a pixel-by-pixel basis. This process created an image where galaxies that are bright at the bluest wavelengths but faint in the NIR (such as low redshift starbursts) are present, but so are galaxies with very faint observer-frame optical fluxes but stronger observer-frame NIR flux (such as faint quiescent galaxies or galaxies at higher redshifts). The multiwavelength detection image maximized the number of galaxies present in the image, and limited any bias in choosing which wavelength to select pixels from. We included in our pixel-by-pixel catalog any pixels that had a $S/N$ ratio greater than five in the detection image.

We labeled galaxies and assigned redshifts using the spectroscopic and photometric redshift catalogs provided by \citet{Rafelski2015}. The photometric redshifts were derived using 11 bandpasses covering the near ultraviolet (NUV) to NIR from WFC3 observations in the UDF. \citet{Rafelski2015} provide photometric redshifts found using two different codes--BPZ \citep{Benitez2000, Benitez2004, Coe2006} and EAZY \citep{Brammer2008}. We chose to use the BPZ redshifts, which had a slightly smaller scatter when compared to spectroscopic redshifts as measured by the normalized median absolute deviation ($\sigma_{NMAD}=0.028$ for BPZ versus 0.030 for EAZY), and a slightly lower outlier fraction (2.4\% versus 5.9\% for EAZY). Although averaging the two redshift measurements is possible, it runs the risk of increasing the catastrophic outlier fraction, and thus we chose to only use a single set of measurements for the photometric redshifts. The spectroscopic redshifts were compiled from nine different sources and contain robust measurements for 169 galaxies.

We assigned pixels to separate galaxies by converting the RA and DEC given in the \citet{Rafelski2015} catalog to pixel coordinates, and used these as initial seeds for basin markers for a watershed segmentation \citep{Soille1990}. Whereas traditional segmentation through deblending with, e.g., SExtractor \citep{Bertin1996}, requires multiple run-throughs with different parameters in order to accurately segment both galaxies that are blended together and large galaxies that should not be split into separate objects \citep[see][]{Barden2012, Rafelski2015}, the watershed algorithm for segmentation can be done straight-forwardly when one already knows roughly where objects are located. Briefly, one can imagine the light profile from our detection image flipped upside down to create ``basins'' which are labeled by the initial markers and bounded by our $S/N$ acceptance criteria. The basins are flooded with water starting from the initial seeds, and where one basin overflows into another marks the delineation between the two objects. The use of initial markers is important for the success of this algorithm; simply using local minima leads to over-segmentation due to noise.

From this segmentation map we created a pixel-by-pixel flux catalog and an unresolved total flux catalog. The fluxes were corrected for extinction due to foreground dust according to \citet{Schlafly2011}. Similarly to \citet{Sorba2015}, in order to compare the same light with the same light, the unresolved flux catalog was made by summing all of the flux from each pixel for each galaxy. We removed any objects flagged as stars in \citet{Pirzkal2005} and removed by eye any objects too close to the borders of the image, or with obvious artifacts. In total, the analysis in the following sections included 162 galaxies with spectroscopic redshifts and 1060 with photometric redshifts. The minimum, median, and maximum number of pixels in the spectroscopic sample were 72, 921, and 17499 pixels respectively. For the photometric sample, we restricted the minimum number of pixels required in order to be included in the analysis to 64. Below this threshold, we found that uncertainties in mass estimates (and specifically the ratio of the unresolved to resolved masses) became inordinately large. The median and maximum number of pixels for the photometric redshift galaxies were 182 and 12228 pixels respectively.  

\subsection{Models}
\label{pxp2:models}

We created model SED templates using the Flexible Stellar Population Synthesis \citep[FSPS;][]{Conroy2009, Conroy2010} code. The model set assumed solar metallicity, a \citet{Chabrier2003} IMF, and a \citet{Calzetti2000} dust law, and were corrected for extinction due to intergalactic hydrogen following the \citet{Madau1995} prescription. Nebular line emission was also included in these model templates \citep{Byler2016}, which is a relatively new feature available to FSPS.

While the assumption of solar metallicity becomes less accurate at higher redshifts, both \citet{Bolzonella2010} and \citet{Pforr2012} found that the differences in mass estimation between different metallicity setups is small. Since in this work we are most interested in recovering stellar masses, fitting with solar metallicity should be sufficient, and helps to reduce the parameter space.

Departing from \citet{Sorba2015}, we chose to parameterize the SFH of our model SEDs as an exponential decay (\ie $\tau$-models), in part for easy comparison with other pixel-by-pixel works that used an exponentially declining SFH \citep{Wuyts2012, Welikala2011}. Whereas our previous work found that photometry of nearby galaxies in the Sloan Digital Sky Survey was well matched by two-component burst SFH models, we found the burst SFH models were poor representations for higher redshift galaxies (further details of how pixel-by-pixel studies are affected by assumed SFH and model grid spacing are located in the Appendix).

Our model grid consisted of log(ages/yr) ranging from 7.7 to 10.1 in steps of 0.05, a dust parameter ranging from 0 to 0.7 in steps of 0.1, and 20 different $\tau$ values spaced roughly evenly logarithmically between 0.3 and 10. Our minimum $\tau$ was chosen based on the recommendation from \citet{Wuyts2011}, who found that forcing the $e$-folding timescale to be greater than several 100 Myr was most effective at reproducing UV+IR measured SFRs. 

\begin{figure*}
  \includegraphics[width=\textwidth]{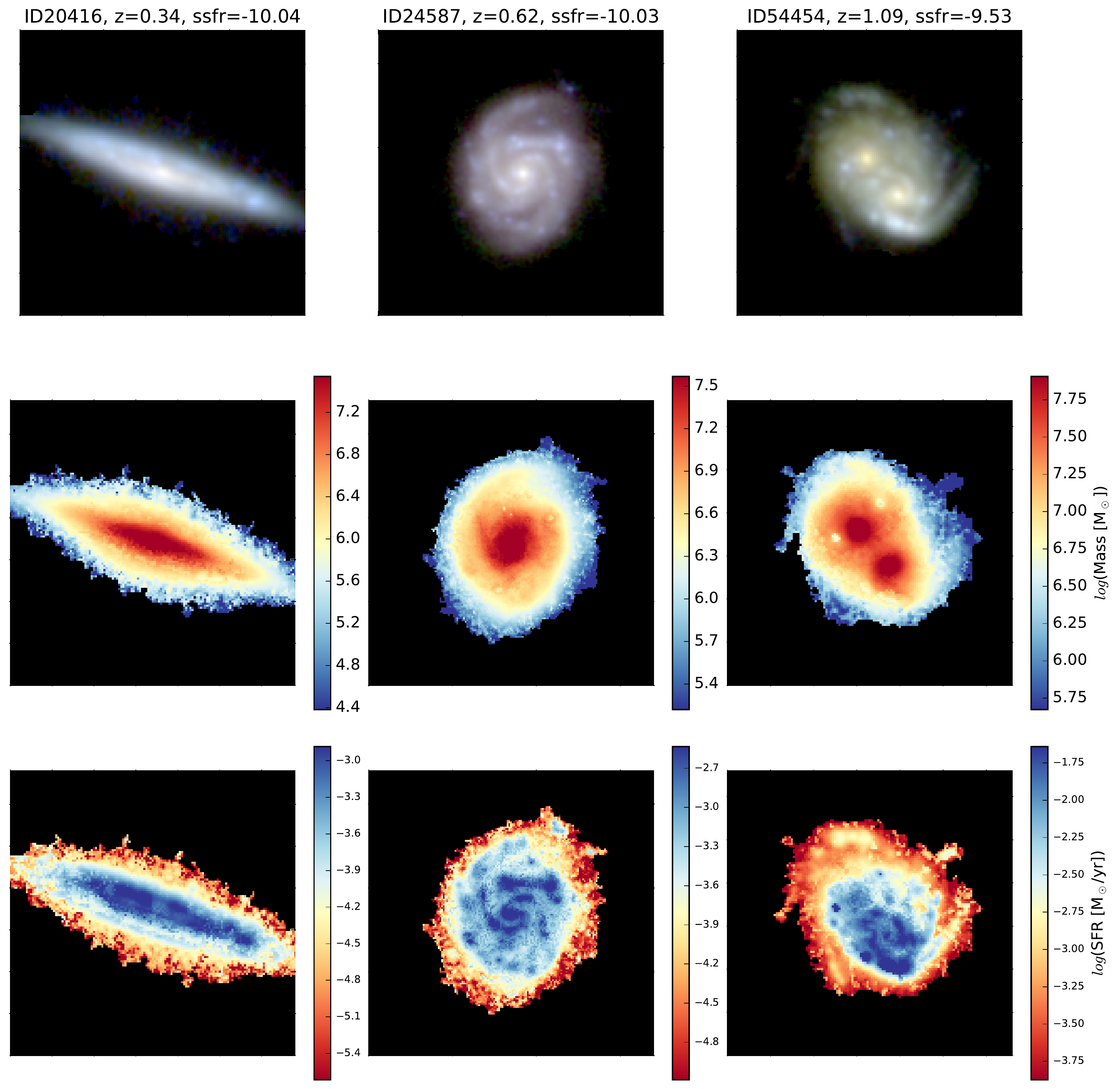}
  \caption{{Two-dimensional maps of three representative XDF galaxies created using our model SED template set. From top to bottom we show false-color images in approximately rest-frame $ugr$, logarithmic stellar mass maps, and logarithmic SFR maps. The galaxies have, from left to right, $RA = \{53.1712061^{\circ}, 53.1699419^{\circ}, 53.14785833^{\circ}\}$, $Dec = \{-27.81471020^{\circ}, -27.7710194^{\circ}, -27.77403611^{\circ}\}$, $z_{spec} = \{0.337, 0.622, 1.088\},$ and $\mathrm{log}{(M_*/\Msun)} = \{10.3, 10.7, 10.9\}$ where the mass measurement is the median pixel-by-pixel mass from our final catalog.}}
  \label{fig:ftaumap}
\end{figure*}

For each galaxy's pixel, we shifted the model spectrum to the redshift given by \citet{Rafelski2015}, defaulting to the spectroscopic redshift when available. Model broadband fluxes were created from each spectrum by convolving with broadband filter transmission curves, and the best-fitting model was found using the SEDfit software package \citep{Sawicki2012a}, which finds the minimum $\chi^2$ in our model parameter space. {\color{black} Fits were restricted to agree with the age of the universe at the given redshift. Example mass and SFR maps are shown in Figure \ref{fig:ftaumap} for three different galaxies. The 2D maps are smooth and lack artificial discontinuities, providing confidence that our model grid is finely sampled enough to account for the range of colours observed. We reiterate that whether or not the models necessarily provide the most accurate mass or SFR measurements is irrelevant for this work due to its comparative nature. However, we have made every effort to follow current best-practices, and our stellar property estimates should be within systematic uncertainty of other works (see \citet{Conroy2013} for a review of parameter estimations from SEDs).}

We use SEDfit to determine uncertainties {\color{black}by constructing probability density functions (PDFs) via several Monte Carlo (MC) realizations}. For each MC realization, the observed flux in each bandpass is changed by adding a normally distributed perturbation (with standard deviation equal to the uncertainty in the observed bandpass and centered on the original measurement). We performed 300 of these realizations for each pixel and used the the median mass estimate to avoid {\color{black}quirks} arising from minimal $\chi^2$ best-fits \citep{Kauffmann2003, Taylor2011}. {\color{black}Using the PDF to study the mass (or other fitted parameters) allows for a fuller understanding of the interplay between observations and model space. While the mass of the minimum $\chi^2$ model may lie anywhere within the PDF, the median value is typically very close to the most-probable-value, defined as the peak of the PDF \citep{Davidzon2017}.  }

\section{Results}
\label{pxp2:results}

\subsection{Outshining at high redshift}

\begin{figure*}
  \includegraphics[width=\columnwidth]{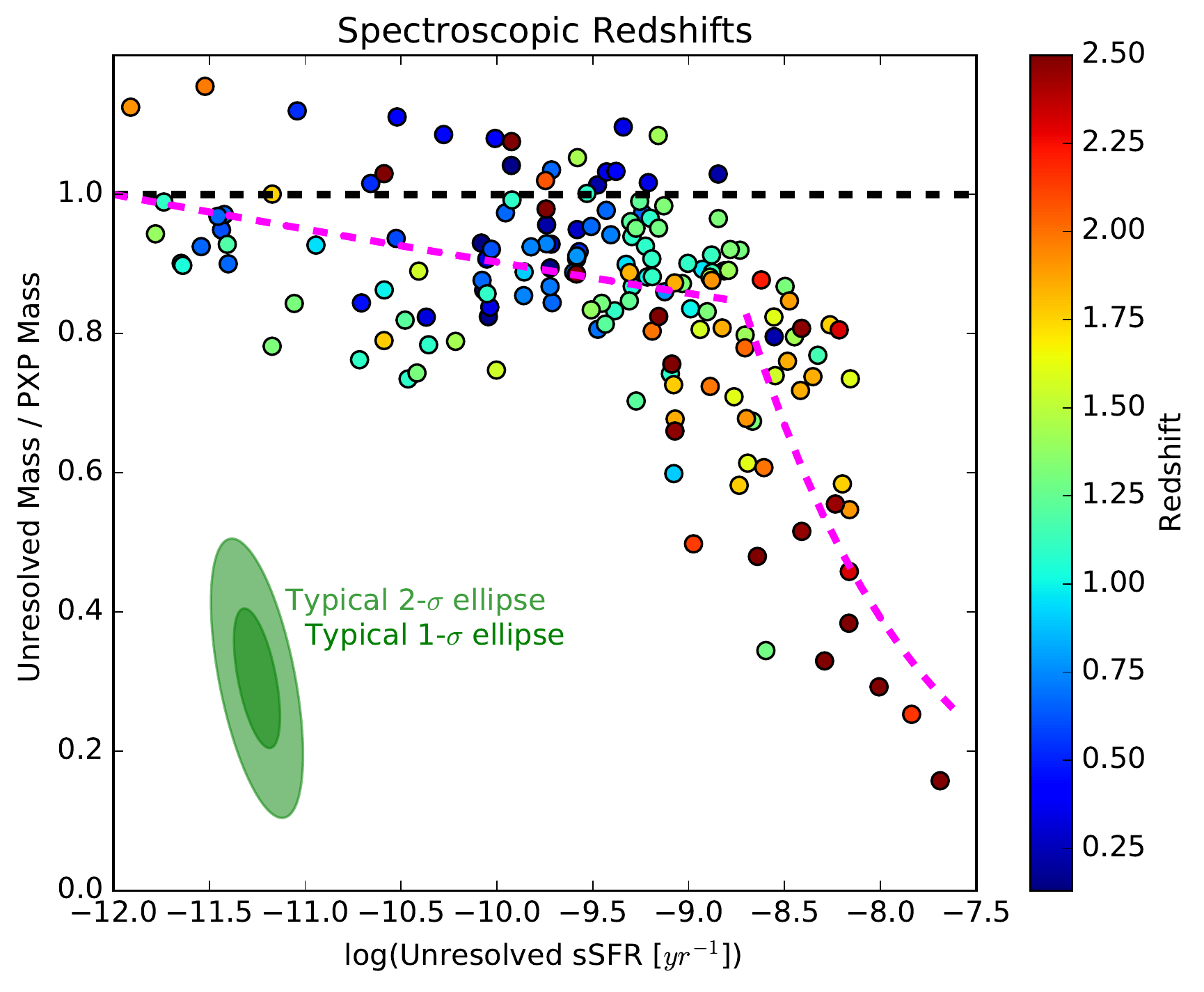}
  \includegraphics[width=\columnwidth]{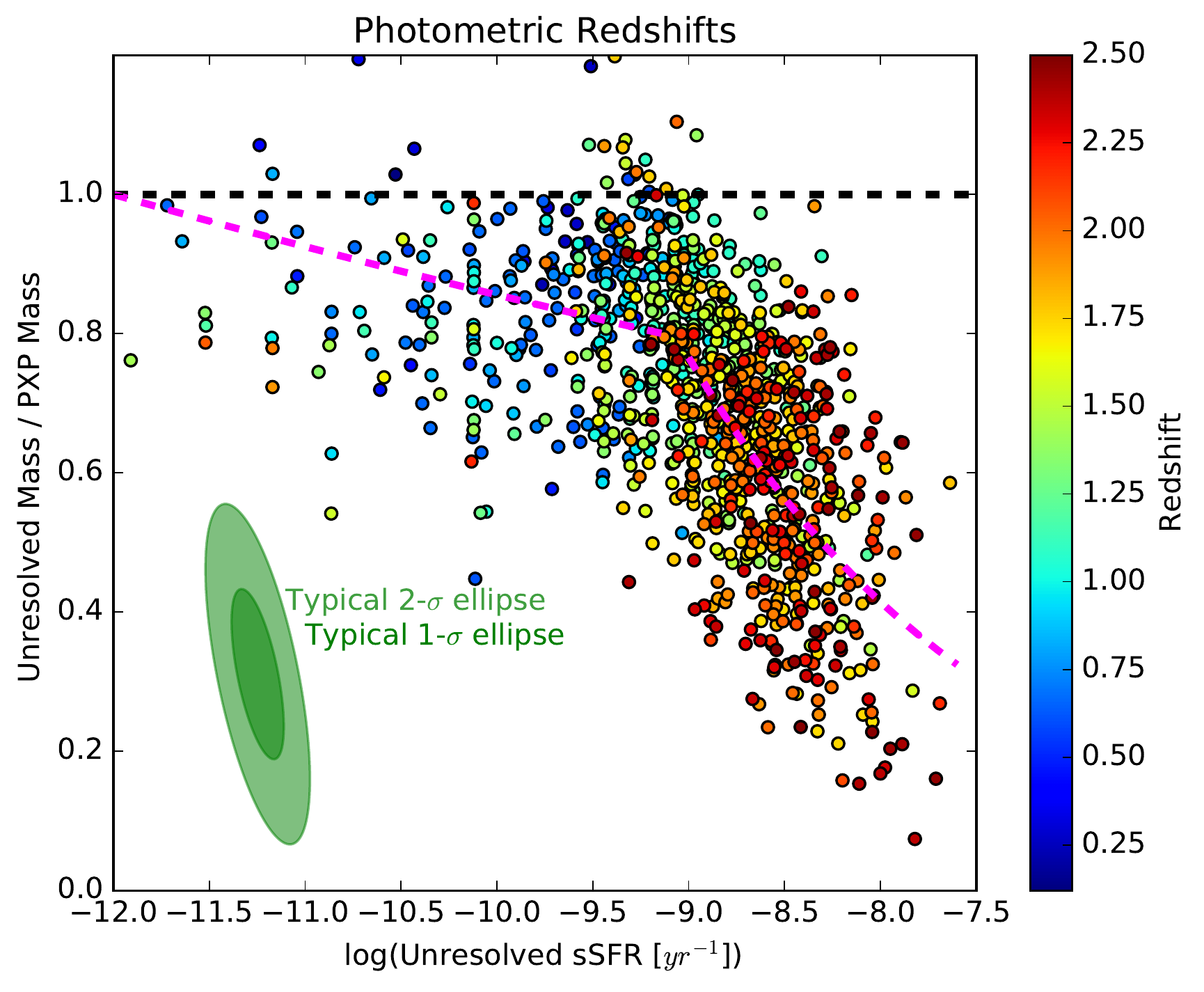}
  \caption{{Ratio of the median unresolved mass estimate and the median pixel-by-pixel mass estimate as a function of specific star formation rate for galaxies in the XDF with spectroscopic redshifts (left) and photometric redshifts (right). Points are colored based on their redshift. The horizontal black dashed line shows where the two mass estimates are equal. The magenta dashed line shows a broken power law fit to the data to demonstrate the two regime behavior.}}
  \label{fig:masscomp}
\end{figure*}

To see if the results of \citet{Sorba2015} extend to galaxies at higher redshift, we plot in the left panel of Figure \ref{fig:masscomp} the ratio of stellar mass estimates derived from unresolved photometry to those determined from pixel-by-pixel SED fitting as a function of specific star formation rate (sSFR) for galaxies with spectroscopic redshifts. The color of each data point is related to the galaxy's redshift (bluer colors for lower redshifts, redder for higher redshifts). If the pixel-by-pixel and unresolved stellar mass estimates were the same, all points would lie along the one-to-one correspondence horizontal dashed black line. For the large majority of galaxies, the mass ratio is below one, meaning that unresolved photometry SED fitting tends to underestimate the true mass, if we assume the pixel-by-pixel mass is closer to the true mass. Looking at the colors of the points, there is no discernible difference in where a galaxy's mass ratio falls as a function of redshift, other than the fact that higher redshift galaxies tend to have higher sSFRs \citep{Feulner2005}. Note that this does not necessarily exclude differing behavior as a function of redshift, only that the large uncertainty in stellar masses obscures any relation if it were present. The typical 1-$\sigma$ and 2-$\sigma$ confidence regions are shown by the dark and light green ellipses {\color{black}and were measured by centering all the MC estimates for each galaxy on the median values and finding the ellipses based on the eigenvectors of the covariance of the point cloud.}  There is a slight correlation introduced by having the unresolved mass present in both axes, but the skew is small compared to the range of data covered, and this correlation could not explain the observable trends in the figure.

\begin{figure*}
  \includegraphics{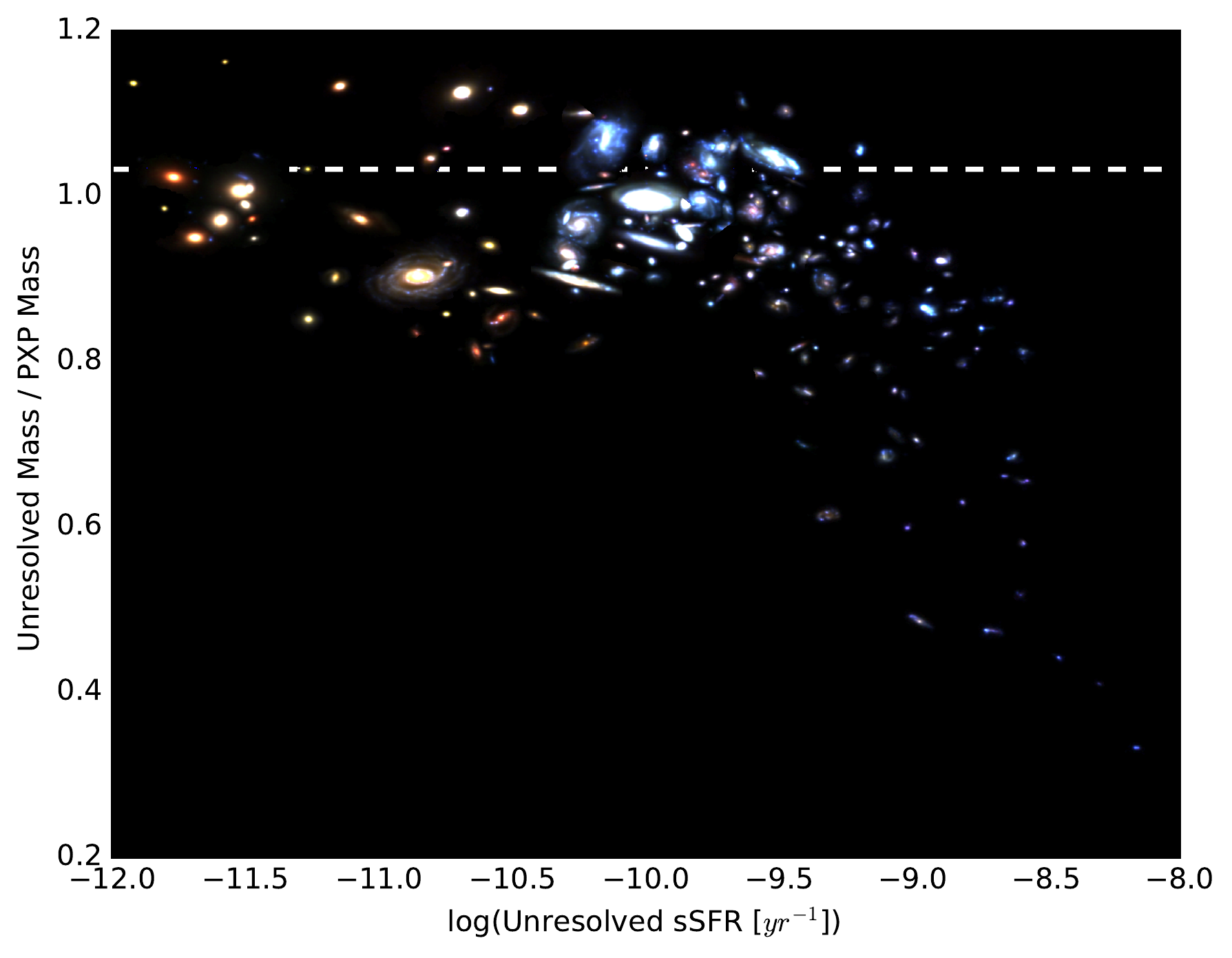}
  \caption{{Ratio of the median unresolved mass estimate and the median pixel-by-pixel mass estimate as a function of specific star formation rate for galaxies in the XDF with spectroscopic redshifts. The false-color cutouts are made such that blue, green, and red correspond to roughly rest-frame $u, g$ and $r$ respectively. No dust lanes or other artifacts seem to be responsible for the large discrepancy at high sSFRs.}}
  \label{fig:ugr}
\end{figure*}

Significantly, there appear to be two regimes in Figure \ref{fig:masscomp}: a roughly linear decrease in the mass ratio at sSFRs less than -8.75, followed by a much sharper decrease beyond. A broken power law fit is shown by the magenta dashed line to demonstrate this dual nature. \citet{Sorba2015} and \citet{Zibetti2009} found that the presence of dust lanes can drastically affect the mass estimates of galaxies compared to their pixel-by-pixel mass estimates, {\color{black}but a visual inspection revealed no obvious dust lanes or other peculiarities present that could explain the mass discrepancy for the most egregious outliers. False color images, as well as mass and SFR maps are provided as an online supplement to this work. For demonstration purposes, the left panel of Figure \ref{fig:masscomp} is also reproduced using false-color images in rest-frame $ugr$ and shown in Figure \ref{fig:ugr}}.

The right panel of Figure \ref{fig:masscomp} a similar effect, but this time using the photometric redshift for galaxies with $z_{phot} < 2.5$ (a limit chosen for ease of comparison with other works; see Section \ref{sec:why}). The same broken power law trend is visible with these galaxies as with the spectroscopic redshift galaxies. Again, no discernible difference in behavior with redshift is distinguishable, but this does not necessarily preclude an offset that steepens with redshift. We chose to focus our analysis on a redshift invariant offset that changes behaviour at a certain sSFR, but it is possible that future surveys with greater galaxy numbers could extract a redshift dependence.

Although the shallow linear decrease at low sSFRs is consistent with \citet{Sorba2015}, the sharper exponential decay at higher sSFRs is surprising, and perhaps alarming. Taken at face value, this would mean that a large number of stellar mass measurements would be underestimated by a factor of 2-3 or some as high as 5 (0.3-0.5 dex or 0.7 dex respectively). The abrupt change in the nature of the mass ratio/sSFR relation hints that there may be other factors involved.

\subsection{Is the effect due to poor signal-to-noise ratio?}

\begin{figure}
  \includegraphics[width=\columnwidth]{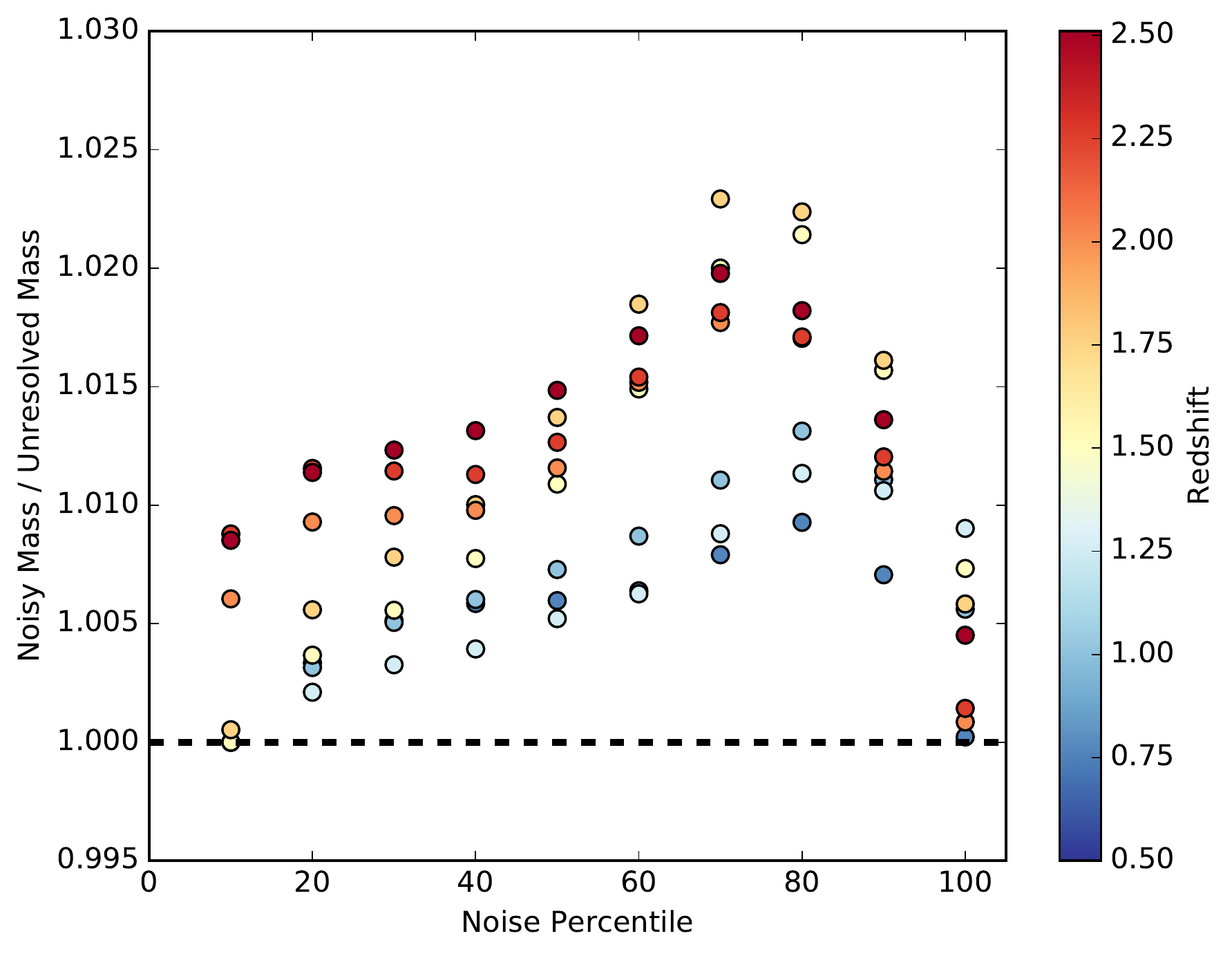}
  \caption{{Ratio of mass estimates from unresolved photometry with artificially increased photometric uncertainties to those with the lowest possible uncertainties. The noise level is set by the $i^{th}$ percentile of the photometric uncertainties of every pixel in every galaxy separated in eight redshift bins from 0.5 to 2.5.}}
  \label{fig:pxp2sntest}
\end{figure}

It is possible that the poorer signal-to-noise ratios for these galaxies is biasing the SED fits, an effect described by \citet{Gallazzi2009}. Indeed if we restrict our examination to only galaxies that have an unresolved median stellar mass estimate greater than $10^{10}$ \Msun\ then the second steeper power law is not visible, but only because these galaxies all have sSFRs lower than the best fitting break sSFR of approximately -9 dex. To see how our mass estimates could be biased by the relatively low signal-to-noise ratios of pixels, we first found how the noise in each bandpass grows, finding the $i^{th}$ percentiles of the photometric magnitude error of every pixel in galaxies split into 8 redshift bins. This yielded the typical magnitude uncertainty (denoted as $\Delta{m}(B,i,z)$), which is inversely proportional to the signal-to-noise and is a function of bandpass($B$), redshift, and ``noisiness'' ($i$),  where $i$ ranges from 10 to 99 in steps of 10 with the $99^{th}$ percentile being the noisiest pixels. We next took the unresolved photometric catalogs and refit them to our model SED grid ten different times, each time replacing the photometric uncertainties with successively noisier and noisier $\Delta{m}$. Figure \ref{fig:pxp2sntest} compares the masses of these noisy fits compared to the mass of the most accurate unresolved mass.

\begin{figure}
  \includegraphics[width=\columnwidth]{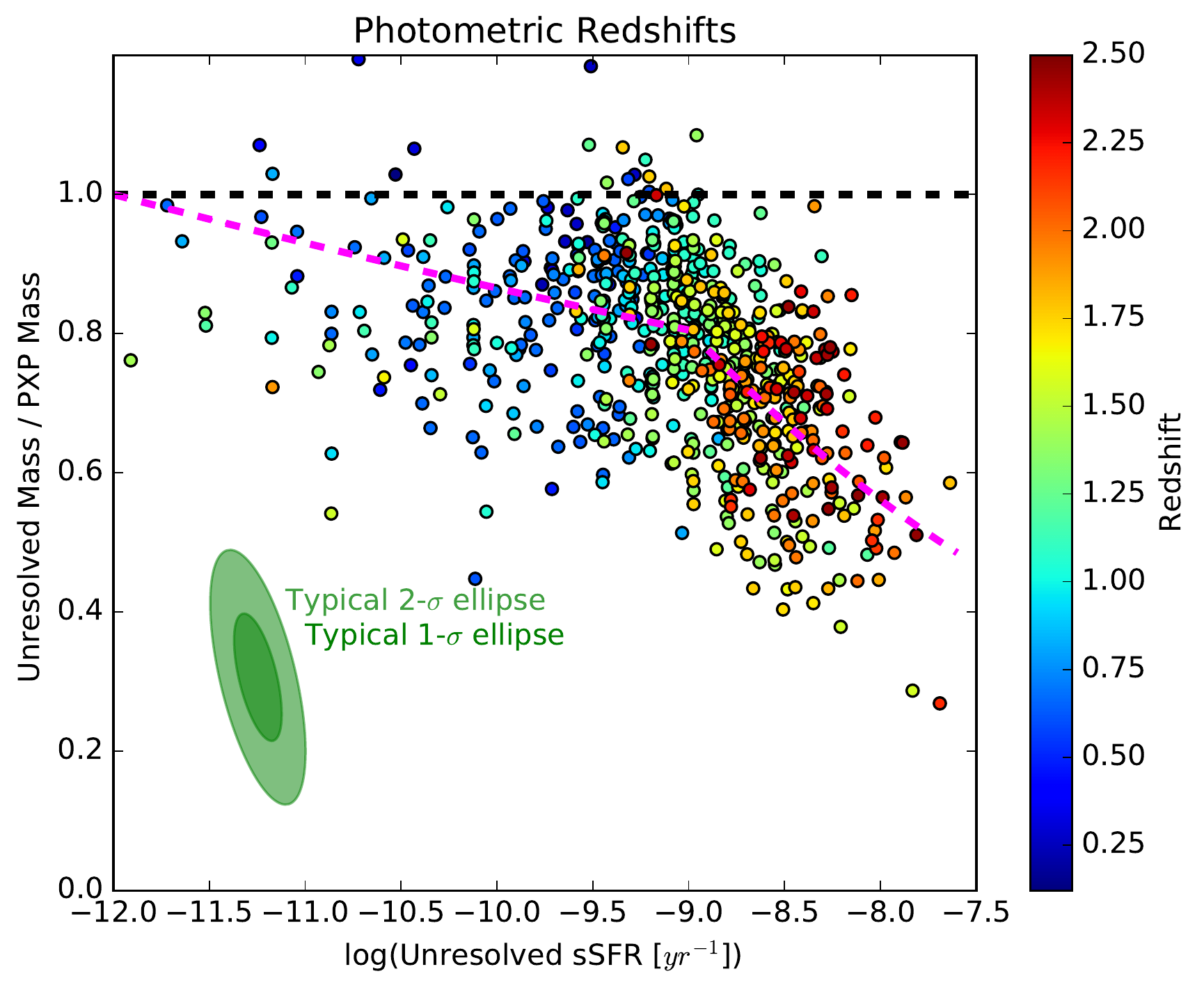}
  \caption{{Same as the right panel of Figure \ref{fig:masscomp}, except removing galaxies whose masses could be unduly biased by noisy pixels. The overall kink at high specific star formation rates in unchanged, which is unsurprising since our maximum estimated bias was only 2.2\%, much lower than could account for large relative mass differences at high sSFRs. }}
  \label{fig:biasCut}
\end{figure}

The increase in mass estimate with increasing noise shows that our model SED set does introduce a prior bias, increasing mass by a maximum of approximately 2.5\% for the 70$^{th}$ noisiest percentile pixels at the higher redshifts. Interestingly, the mass bias begins to decrease as the pixels become even noisier than that. There is a dichotomy between pixels with redshifts above approximately 1.5; although the curves all have a similar qualitative rise and fall, the higher redshift bias increases more steeply. Now, this is a fractional bias, and the noisiest pixels will also in general be the least massive, and thus contribute less to the total mass bias of the galaxy. For each galaxy, we calculate a predicted fractional mass bias ($F$) by
\begin{equation}
  F = \frac{\sum_x M^*_x B_{p(x), z} - \sum_x M^*_x}{\sum_x M^*_x}
\end{equation}
where $M^*_x$ is the median stellar mass in the $x^{th}$ pixel and $B_{p(x), z}$ is the bias interpolated from Figure \ref{fig:pxp2sntest} where $p(x)$ is the average photometric error percentile in that pixel and $z$ is the galaxy's redshift. The maximum mass bias from low S/N pixels was 2.2\% of the mass of the galaxy, which is not significant compared to the factor of two (or more) discrepancy. For demonstration purposes, we removed from our sample any galaxies with predicted fractional mass biases greater than 1\%. Doing so left us with 684 galaxies shown in Figure \ref{fig:biasCut}, and shifted the median number of pixels in these galaxies to 242 pixels. The double power law is still present, and the feature remains even with more stringent cuts of $F$ < 0.005. {\color{black} We conclude that this effect is not due to poor $S/N$.}

\subsection{Is the effect due to poor lack of redder bandpasses?}

The sharp downturn happens almost exclusively for high redshift ($z > 1.5$) galaxies, where our reddest bandpass ($F160W$) is only probing rest-frame visual regions of the galaxies' spectra. A large number of works \citep{Maraston2006, Kannappan2007, Bolzonella2010, Lee2009, Ilbert2010, Pforr2012} found that coverage in the rest-frame NIR was crucial to stellar mass estimates from SED fitting, with \citet{Pforr2012} stating that excluding NIR bandpasses can lead to masses that are underestimated by up to 2 dex. However, \citet{Shapley2005}, \citet{VanderWel2006}, and \citet{Muzzin2009} did not find as strong a need for NIR coverage. Moreover, our pixel-by-pixel versus unresolved comparison should have both the pixel-by-pixel and unresolved masses affected in a similar manner by the lack of NIR data. It should be emphasized that our method examines relative mass differences, and is unconcerned with evaluating the true total stellar mass in each galaxy. 

\begin{figure}
  \includegraphics[width=\columnwidth]{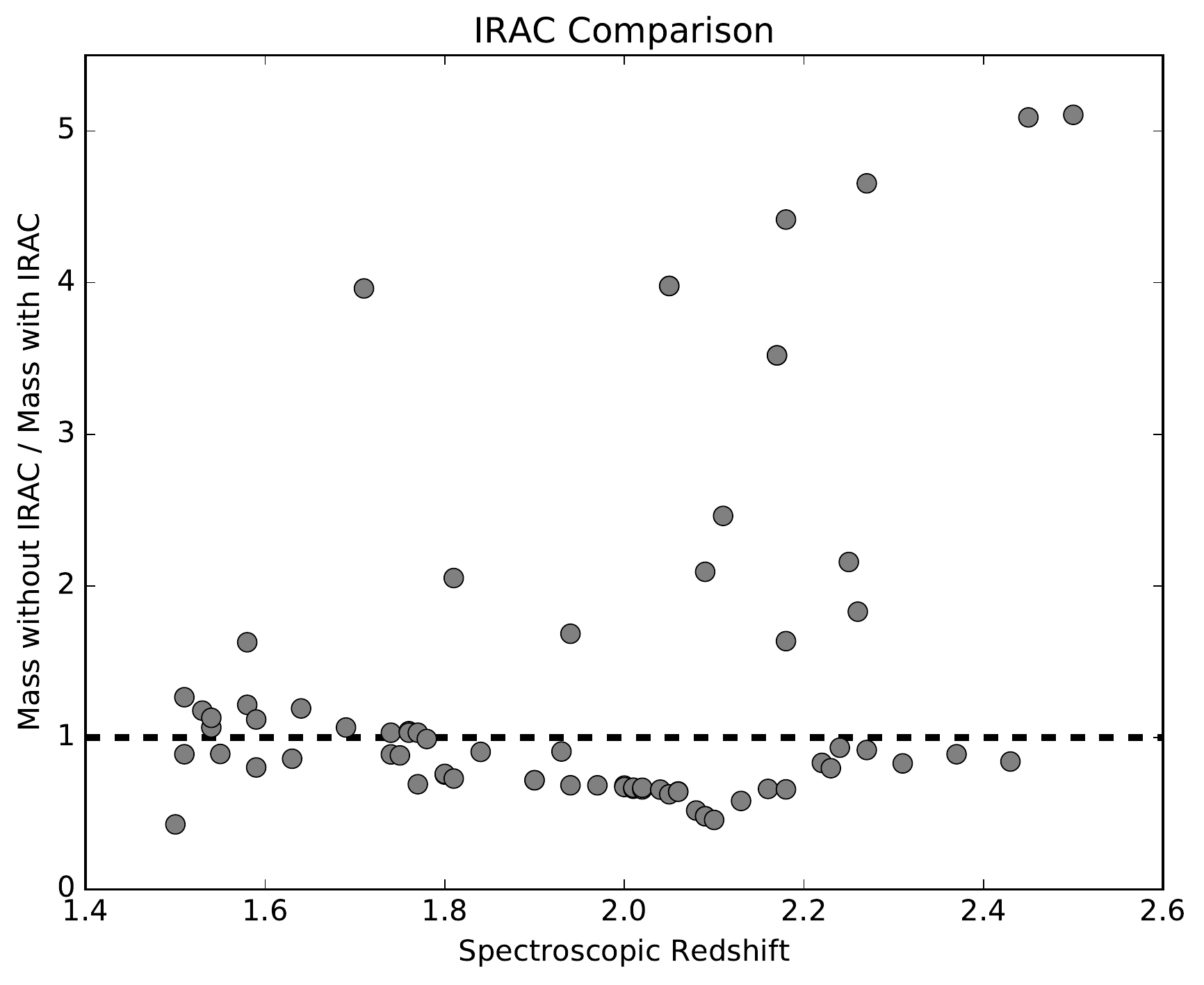}
  \caption{{Comparison of mass estimates using photometry from \citet{Rafelski2015} and either including or excluding IRAC photometry from \citet{Lundgren2014}. }}
  \label{fig:iraccomp}
\end{figure}

To see if our lack of redder bandpasses for high redshift galaxies could be affecting our mass estimates, we make use of the photometric catalogs provided by \citet{Lundgren2014}, who included Spitzer/IRAC photometry. We cross-correlated that catalog with that of \citet{Rafelski2015} based on a search in RA and DEC utilizing a maximum matching distance of 3 arcseconds (approximately five pixels in the XDF images), matching only galaxies that had IRAC 3.6 and 4.5 measurements and with a photometric redshift between 1.5 and 2.5. To ensure that the objects in each catalog were in fact representing the same galaxy, we compared the $F606W$ flux in each catalog and removed any objects where this did not match to within the 1-sigma uncertainties, leaving 71 galaxies. We performed (unresolved) SED fitting on these galaxies once using only photometry from the \citet{Rafelski2015} catalog in the nine XDF bandpasses, and a second time also including the two IRAC bandpasses from \citet{Lundgren2014}. As shown in Figure \ref{fig:iraccomp}, we found no consistent bias in the median MC mass estimates of these $z \sim 2$ galaxies when either including or excluding the two IRAC bandpasses. The median mass ratio ($M^*_{no IRAC} / M^*_{with IRAC}$) of the 71 galaxies was 0.89 with upper and lower inner quartile bounds at 1.24 and 0.68 respectively. Examining solely galaxies with $1.9 < z < 2.2$, the median mass ratio does drop to 0.68, leading one to think the effect may become more dramatic as redshift increases, but then the median mass ratio rises up to 0.93 again for galaxies with $z > 2.2$, suggesting that the previous dip was likely noise. Although the masses were, in general, slightly smaller without IRAC photometry, the discrepancy was nowhere near as drastic as that found in the pixel-by-pixel analysis. There were also no discernible trends with either sSFR or redshift. It is further questionable whether any differences in the mass estimates are due to the presence of IRAC photometry, or due simply to systematic differences between the two catalogs used here.

\begin{figure}
  \includegraphics[width=\columnwidth]{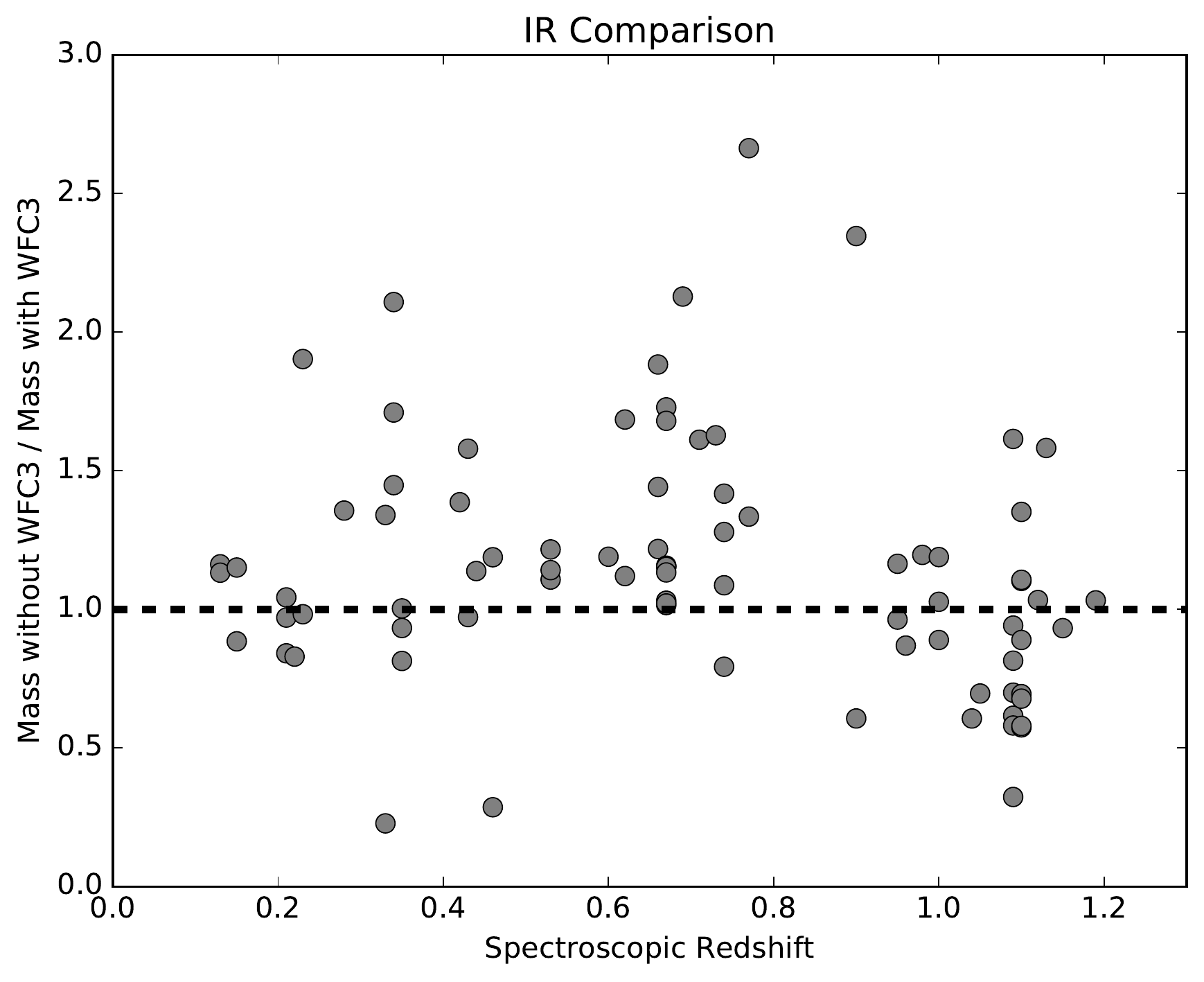}
  \caption{{Comparison of mass estimates for $z_{spec} < 1.2$ galaxies made both including WFC3 photometry and  excluding it.}}
  \label{fig:wfc3comp}
\end{figure}

To more deeply probe the effects of including rest-frame NIR photometry on our mass estimates, we refit galaxies with spectroscopic redshifts less than 1.2, only this time excluding the four WFC3 bandpasses. This removes their rest-frame NIR bands, allowing us to test whether NIR data is important. The results are shown in Figure \ref{fig:wfc3comp}. Similarly to the mass estimates with and without IRAC photometry, the mass estimates of these lower redshift galaxies show no obvious systematic bias when the WFC3 photometry is included/excluded. Again no obvious trend with redshift is present. The median mass ratio when comparing the mass estimates without the four WFC3 bandpasses versus those with is 1.11, with a lower-quartile value of 0.89 and an upper-quartile value of 1.35. Interestingly, this slight offset is in the opposite direction of that found with the IRAC bandpasses. Here, mass estimates including rest-frame NIR photometry are typically lighter than when NIR photometry is excluded.

In either case, even if rest-frame NIR photometry does have an effect on the absolute mass estimate, the relative mass ratio from comparing pixel-by-pixel to integrated mass estimates for any one galaxy should not be altered, since both mass estimates are made using the same set of bandpasses. {\color{black} Despite the lack of influence from including rest-frame NIR observations, the large body of work stating their significance has led us to divide our analysis into two parts.} In the next section (\ref{sec:z1.2}), we restrict our analysis to galaxies with redshifts less than 1.2, meaning all galaxies have photometric observations up to at least rest-frame $I$ band, and compare to previous results from \citet{Sorba2015}. In Section \ref{sec:z2.5}, we examine the increasing discrepancy for galaxies up to $z = 2.5$, and discuss the implications if the effect is real.

\subsection{Outshining up to redshift 1.2}
\label{sec:z1.2}

\begin{figure}
  \includegraphics[width=\columnwidth]{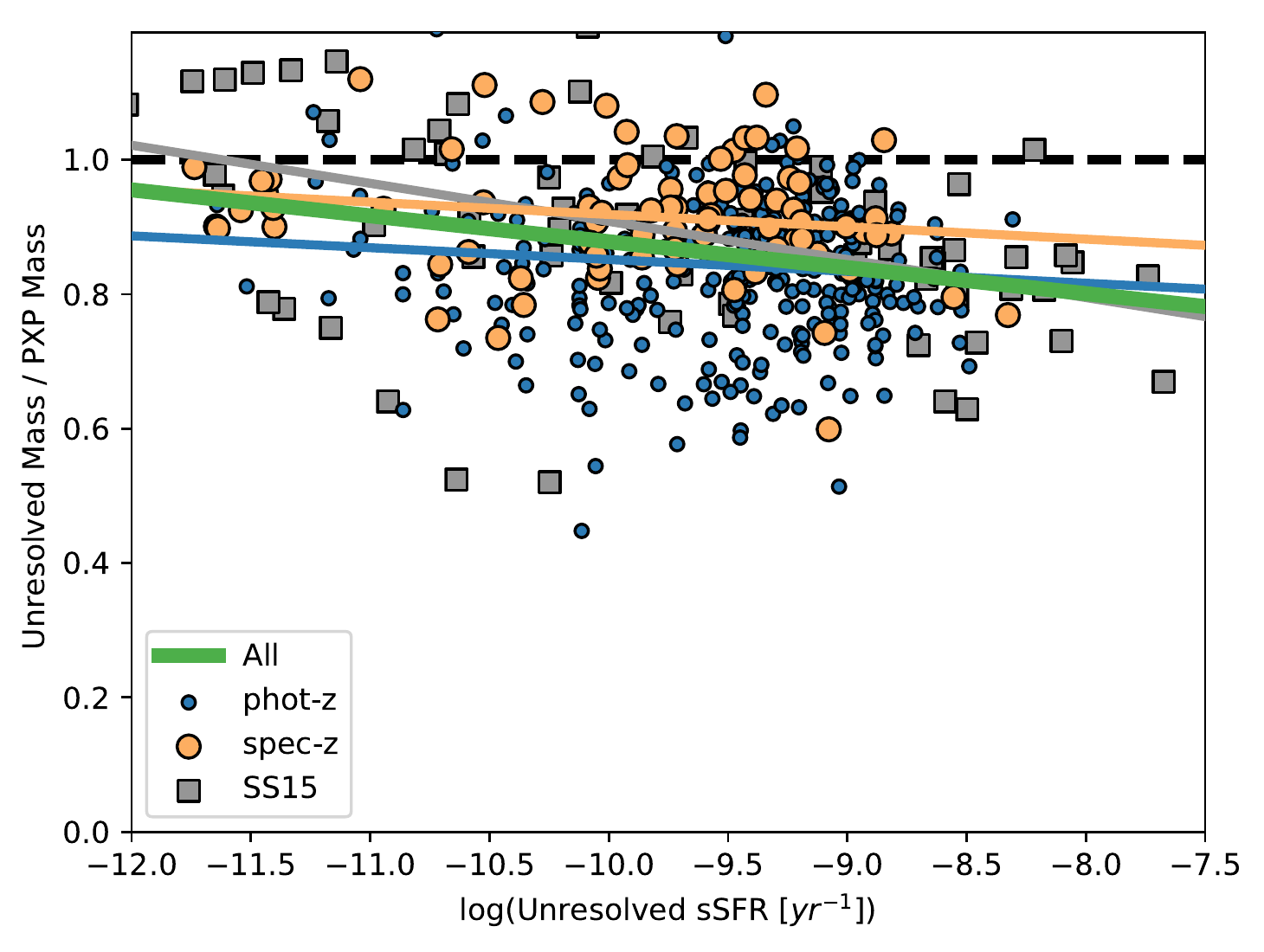}
  \caption{{Linear fits to the resolved pixel-by-pixel mass ratio as a function of sSFR. Yellow (blue) points represent XDF galaxies with spectroscopic (photometric) redshifts less than 1.2. The grey squares are $z \sim 0$ SDSS galaxies from \citet{Sorba2015}. The yellow, blue, and grey colored lines show linear fits to their corresponding dataset. The thick green line shows a linear fit to the combined datasets. It represents our new best relation for correcting outshining for galaxies in the range $0 < z < 1.2$. }}
  \label{fig:bothmasscomp}
\end{figure}

Figure \ref{fig:bothmasscomp} shows the XDF $z_{spec} \leq 1.2$ and $z_{phot} \leq 1.2$ samples as yellow and blue circles respectively, and the SDSS galaxies from \citet{Sorba2015} as gray squares. 
The best-fit parameters can be found in Table \ref{tab:linearfits}, and uncertainties were found by refitting lines for each of our MC instances and taking the 16th and 84th percentile values. The slopes and intercepts of all three datasets agree to within uncertainty. The fact that the $z \sim 0$ sample falls so well along the same relation as the XDF samples despite differences in photometry and modeling strongly implies that any effects of outshining do not change significantly up to $z \sim 1$. However, there may be a slight steepening of the slope as redshift decreases from approximately one to the present day, accompanied by a subsequent rise in the intercept. The best-fit slope of the XDF spec-z and photo-z samples together gives a slope of -0.027 which is shallower than the best-fit SDSS slope of -0.057, and the slope fit to just the spectroscopic XDF sample is shallower still at -0.017. There is a great deal of scatter in the data about the linear trend. Partly, this is due to the effect of dust lanes on pixel-by-pixel fitting \citep{Zibetti2009, Sorba2015}, and partly due to the inherent noise in stellar mass estimates. Additionally, the linear fits to the XDF data may be affected by the artificial cut-off at $z = 1.2$, as well as the relative lack of low sSFR galaxies at higher redshifts. Combining the XDF and \citet{Sorba2015} galaxies, we fit an updated linear trend shown as the solid red line in Figure \ref{fig:bothmasscomp}. This represents our new best estimate at a correction factor for outshining:
\begin{equation}
  M^*_{resolved} = \frac{M^*_{unresolved}}{-0.039\log(sSFR) + 0.50}
\end{equation}
where the $M^*$ are given in \Msun\ and sSFR is in units of $yr^{-1}$.

\begin{center}
\begin{table}

\centering
\caption{\label{tab:linearfits} Best fit parameters of linear fits $y=mx + b$. All galaxies have $0 < z \leq 1.2$. Here, $z_s$ and $z_p$ respectively refer to spectroscopic and photometric redshift samples.}
\begin{tabular}{@{}lcc}
\hline
Data Set & $m$ & $b$ \\
\hline
SS15 & $-0.057^{+0.022}_{-0.019}$ & $0.34^{+0.20}_{-0.18}$ \\
XDF $z_s$ & $-0.017^{+0.008}_{-0.021}$ & $0.75^{+0.07}_{-0.22}$ \\
XDF $z_p$ & $-0.019^{+0.005}_{-0.012}$ & $0.66^{+0.05}_{-0.11}$ \\
XDF ($z_s + z_p$) & $-0.027^{+0.007}_{-0.007}$ & $0.60^{+0.07}_{-0.06}$ \\
XDF ($z_s + z_p$) + SS15 & $-0.039^{+0.010}_{-0.008}$ & $0.48^{+0.10}_{-0.07}$ \\
\hline
\end{tabular}
%\end{centering}
\end{table}
\end{center}

\subsection{Outshining up to redshift 2.5}
\label{sec:z2.5}
\begin{figure}
  \includegraphics[width=\columnwidth]{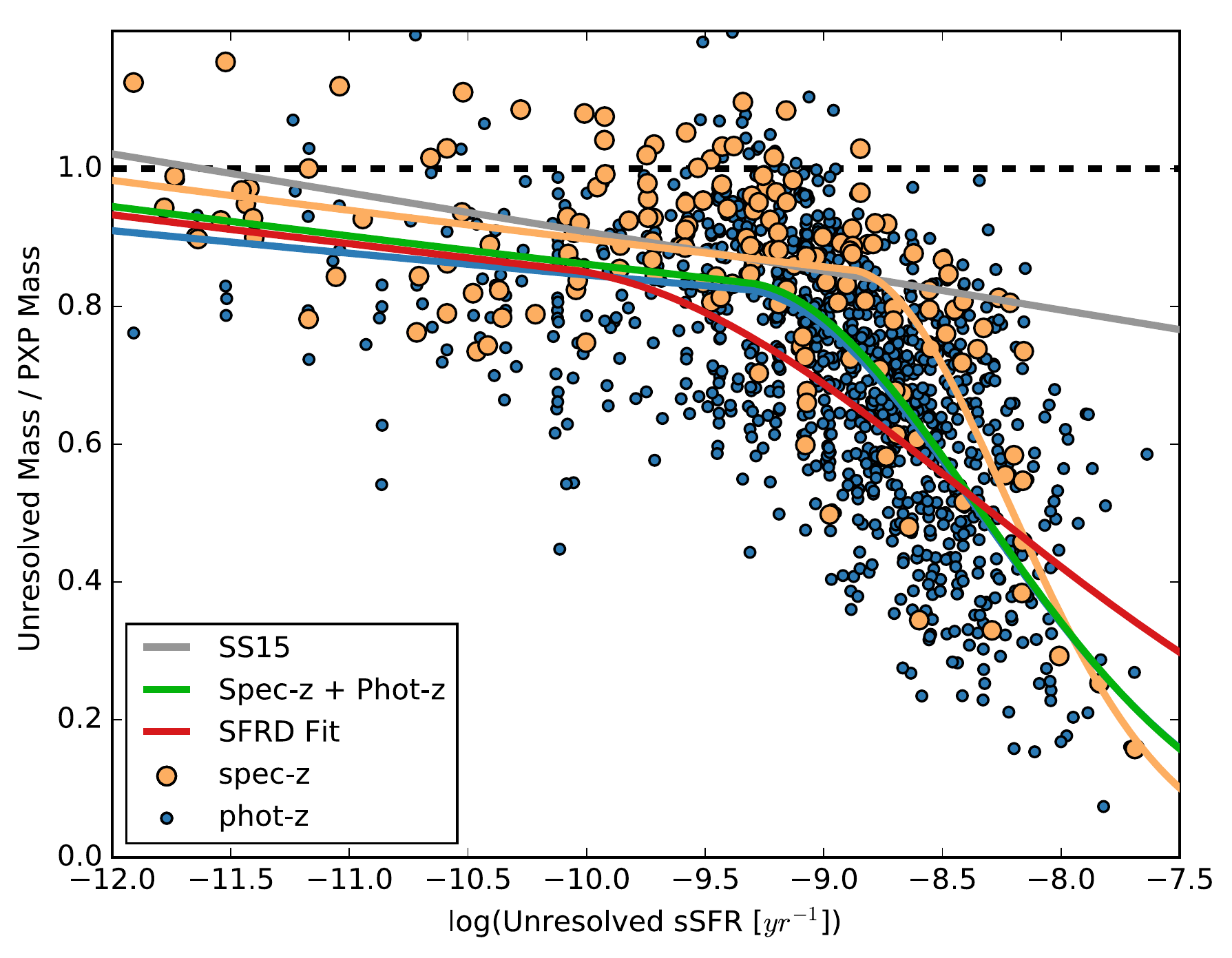}
  \caption{{%Same as Figure \ref{fig:hizmasscomp}, but including all galaxies.
  Same as Figure \ref{fig:bothmasscomp}, but including galaxies up to redshift 2.5. Piecewise linear-parabolic fits in log-log space are shown in yellow (fit to the spectroscopic redshift sample) and green (fit to both the spectroscopic and photometric redshift samples). The linear relation from \citet{Sorba2015} is shown for comparison in grey, but is not included in any of the piecewise fits. The red line shows the correction that best matches the observed SFRD history from \citet{Madau2014} (see Section \ref{sec:missingMass}).}}
  \label{fig:hizmasscompwithnoise}
\end{figure}

For galaxies at higher redshifts (and thus generally higher sSFRs), the unresolved versus pixel-by-pixel mass ratio decreases sharply at sSFRs greater than roughly $10^{-9}$ yr$^{-1}$. In Figure \ref{fig:hizmasscompwithnoise}, we show a piecewise fit to this behavior for galaxies at both low and high redshifts in our sample.
The piecewise function is described in log-log space as
\begin{equation}\label{eq:piecewiseFit}
f(x) = \left\{
\begin{array}{lr}
\mu(x+12) + \beta & : x \leq p\\
k_2x^2 + k_1x + k_0 & : x > p,
\end{array}
\right.
\end{equation}
which is a linear relation to the left of the break point $p$, and a parabola to the right. The free parameters are $p$, $\mu$, $\beta$, and $k_2$. To ensure continuity  and differentiability at $p$, $k_0$ and $k_1$ are constrained to be
%\begin{equation}
  \begin{align}
    k_1 &= -2k_2p + \mu\\
    k_0 &=  \beta + 12\mu + (\mu - k_1)p - k_2p^2.
  \end{align}
%\end{equation}
Because of the nature of a piecewise function that can have a shallow slope abruptly changing to a extremely steep slope, we use orthogonal distance regression (ODR) to find the best fitting parameters, listed in Table \ref{tab:hizparams}. ODR minimizes the orthogonal distance from the fitted line to each point (\ie the line going through the point that intersects the fitted line at a right angle), rather than minimizing the vertical distance from each point to the fitted line as is done in ordinary least squares minimization. In this manner, galaxies which are close to the break point $p$ are matched to the part of the piecewise function they are nearest to, rather than whichever part they happen to be above or below. Uncertainties were again found by performing the same fitting procedure on each of the Monte Carlo instances and taking the 16th and 84th percentiles for each parameter.

\begin{center}
\begin{table*}
\centering
\caption{\label{tab:hizparams} Best fit parameters of piecewise fits. Here, $z_s$ and $z_p$ respectively represent the spectroscopic and photometric XDF samples.}
\begin{tabular}{@{}lccccc}
\hline
Data Set & $p$ & $\mu$ & $\beta$ & $k_2$\\
\hline
XDF $z_s$ & $-8.88^{+0.04}_{-0.18}$ & $-0.020^{+0.010}_{-0.004}$ & $-0.007^{+0.010}_{-0.018}$ & $-0.47^{+0.14}_{-0.05}$\\
XDF $z_p$ & $-9.34^{+0.49}_{-0.28}$ & $-0.016^{+0.006}_{-0.008}$ & $-0.041^{+0.043}_{-0.016}$ & $-0.20^{+0.12}_{-0.31}$\\
XDF ($z_s + z_p$) & $-9.32^{+0.47}_{-0.26}$ & $-0.020^{+0.011}_{-0.004}$ & $-0.024^{+0.027}_{-0.001}$ & $-0.21^{+0.12}_{-0.31}$\\
\hline
\end{tabular}
\end{table*} 
\end{center}

\section{Discussion}
\label{pxp2:Discussion}

\subsection{Implications}
\label{pxp2:imp}

\subsubsection{Corrections to Stellar Mass Measurements}

{\color{black} In Section~\ref{pxp2:results} we have shown that spatially-resolved SED-fitting can give significantly higher stellar mass estimates than unresolved photometry.  Taking the spatially-resolved fits as more accurate than the cruder, spatially-unresolved ones, this suggests that stellar mass estimates done using spatially-unresolved photometry need to be corrected upwards. We recommend that those working with stellar masses derived from integrated photometry apply the empirical correction based on  Eq.~\ref{eq:piecewiseFit} with constants derived for the spectroscopic+photometric case (third row in Table~\ref{tab:hizparams}).  The correction then is:
\begin{equation}\label{eq:massCorrection}
M^*_{corrected}
= M^*_{unresolved}\times \left\{
\begin{array}{lr}
10^{0.02s\ +\ 0.264} & : s \leq -9.23\\
10^{0.21s^2\ +\ 3.9344s\ +\ 18.505} & : s > -9.23,
\end{array}
\right.
\end{equation}
where $s$ is the logarithm of the sSFR in units of yr$^{-1}$ and the $M^*$ quantities are in solar masses.

In \S~\ref{sec:slopeOfMS}--\ref{sec:missingMass} we apply this correction to three concrete situations. 
}

\subsubsection{Slope of the Main Sequence}\label{sec:slopeOfMS}

\begin{figure}
  \includegraphics[width=\columnwidth]{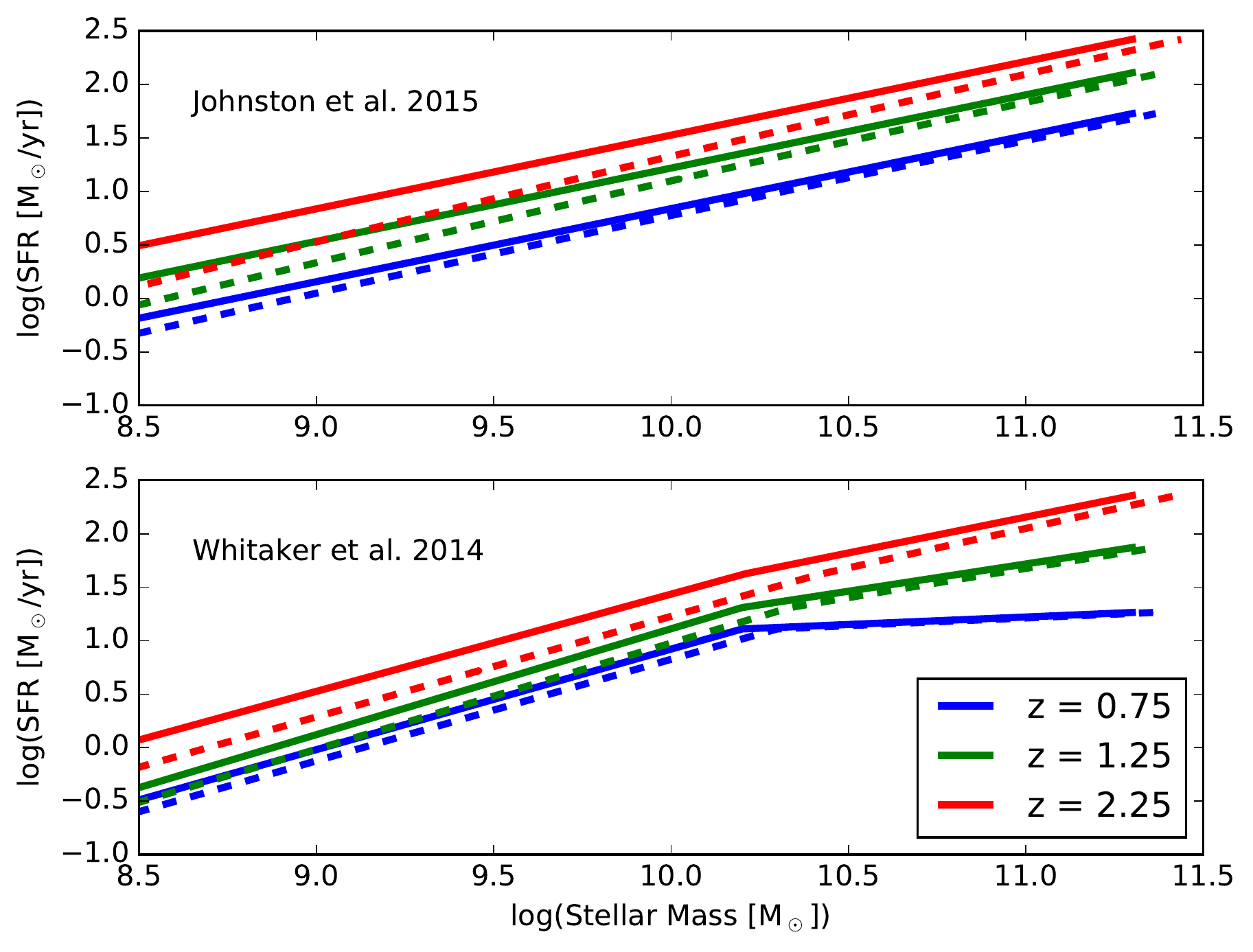}
  \caption{{Corrections to previously published star-forming main sequence relations at various redshifts made using our piecewise fit to the XDF ($z_s + z_p$) galaxies. The top panel shows relations from \citet{Johnston2015} and the bottom panel those of \citet{Whitaker2014}. The solid lines show the original relationships, and the dashed lines the corrections based on this work. The blue, green, and red colors show redshift 0.75, 1.25, and 2.25 respectively.}}
  \label{fig:sfms}
\end{figure}

The increasing disparity between unresolved and pixel-by-pixel mass estimates, if real, has implications for several previously published results derived using masses from SED template fitting. Perhaps most obviously, the relationship between a galaxy's mass and SFR (when it is actively star-forming), known as the star-forming main sequence, is directly affected. In Figure \ref{fig:sfms} we show how correcting the galaxies' masses for outshining would change two recent parameterizations of the star-forming main sequence from \citet{Whitaker2014} and \citet{Johnston2015} at three different redshifts. The sharp decline at higher sSFRs leads to certain areas of the mass-SFR plane changing more drastically than others. For the \citet{Johnston2015} parameterization, the lowest redshift main sequence is barely affected at all, but there is a much larger correction for the higher redshift galaxies because of their increased SFR. The larger correction at higher sSFRs leads to a main sequence which is no longer linear. Non-linear main sequences were also proposed by \citet{Whitaker2014}, who fit broken power laws to star-forming galaxies in different redshift ranges. The corrections due to outshining are less dramatic for this parameterization because the kink in the main sequence brings it down out of the highest sSFR region of the mass-SFR plane. Once again, the lowest redshift relation is barely affected, but the $z = 2.25$ has a higher offset.

\subsubsection{Intrinsic Scatter in the Main Sequence}\label{sec:scatterOfMS}

\begin{figure}
  \includegraphics[width=\columnwidth]{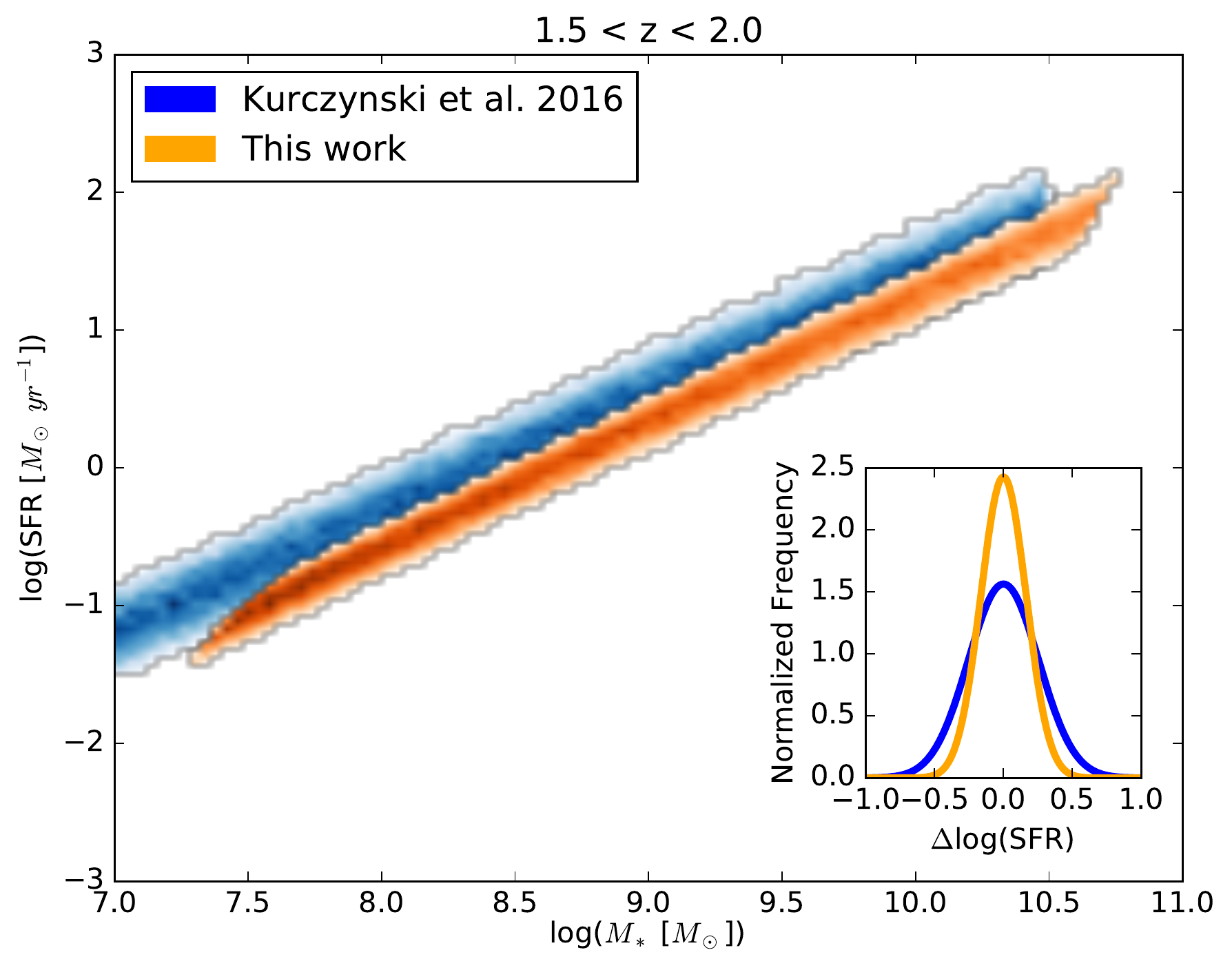}
  \caption{{Effects of our mass correction on the intrinsic scatter about the star-forming main-sequence for a redshift range $1.5 < z < 2.0$. The blue density map is generated from \citet{Kurczynski2016} relations and converted to Chabrier IMF. The orange density map shows the correction to stellar mass based on this work. The shearing caused by the sSFR dependent correction leads to a lower intrinsic scatter in log(SFR) (0.17 versus 0.26), as shown by the distributions about the central relation shown in the inset.}}
  \label{fig:scatter}
\end{figure}

Additionally, the intrinsic scatter of the star-forming main sequence is linked to smoothness (or, conversely, burstiness) of the SFH of the universe \citep{Kurczynski2016, Abramson2014}. The small observed scatter around the SFR-$M_*$ correlation implies a steady, gradual assembly of stellar material rather than a history dominated by bursts. Because the bias in mass estimation found here is linked to a galaxy's sSFR, it causes a shear in the intrinsic scatter. Galaxies located in higher sSFR regions of the SFR-$M_*$ plane are affected more so than others. {\color{black} In Figure \ref{fig:scatter}, we show that the shearing leads to a smaller intrinsic scatter around the SFR-$M_*$ relation for the redshift range $1.5 < z < 2$. The blue density map shows 500,000 galaxies randomly produced using the intrinsic scatter and star-forming main-sequence parameters of \citet{Kurczynski2016} adapted to a Chabrier IMF rather than a Salpeter. These galaxies were corrected for the effects of outshining based on this work, and the result shown as the orange density map. The shearing and subsequent tightening of the correlation can be seen by the standard deviations about the central relation shown in the inset. Although only one redshift range is shown here, a similar effect happens at all redshifts. The results are summarized in Table \ref{tab:scatter}.
The tighter scatter around the star forming main sequence implies that the formation of stars in galaxies is even smoother than previously thought.}

\begin{center}
\begin{table}
\centering
\caption{\label{tab:scatter} Corrections to intrinsic scatter in log(SFR) around SFMS for various redshift ranges. Original standard deviations are from \citet{Kurczynski2016}. }
\begin{tabular}{@{}lcc}
\hline
Redshift & $\sigma_{int}$ & $\sigma_{int\_corrected}$\\
\hline
$0.5 < z < 1.0$ & 0.43 & 0.34\\
$1.0 < z < 1.5$ & 0.27 & 0.21\\
$1.5 < z < 2.0$ & 0.26 & 0.17\\
$2.0 < z < 2.5$ & 0.28 & 0.21\\

\hline
\end{tabular}
\end{table} 
\end{center}

\subsubsection{The Missing Mass Problem}\label{sec:missingMass}

\begin{figure}
  \includegraphics[width=\columnwidth]{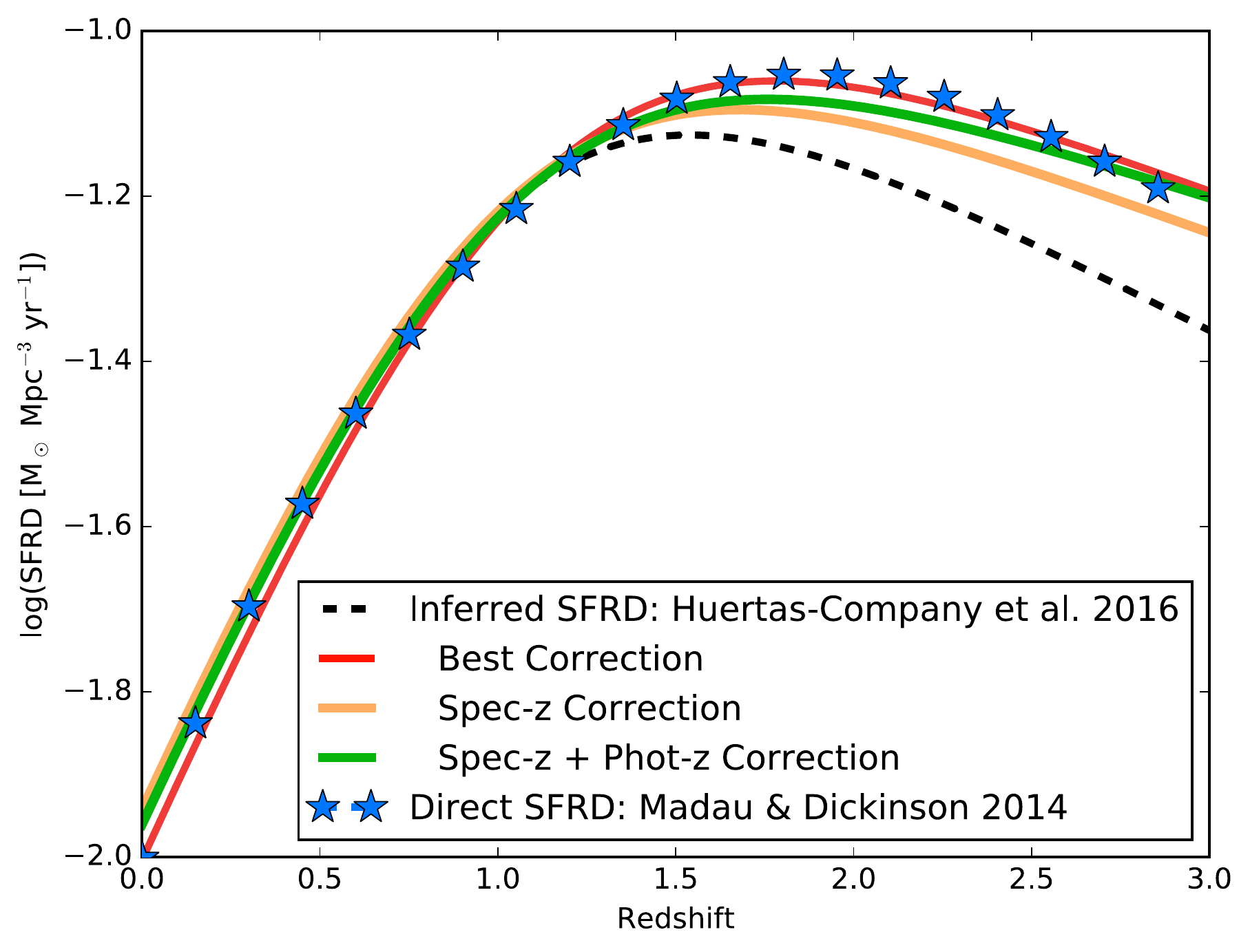}
  \caption{{Corrections to {\color{black} the observed cosmic star-formation history. The dashed black line shows the SFRD curve inferred from the observed stellar mass density (we show the results of \citet{Huertas-Company2016}), and the solid colored lines the corrected values based on our work. The corrections bring the inferred SFRD into much better agreement with the} direct observations of the SFRD compiled by \citet{Madau2014}, shown as the blue stars}}
  \label{fig:sfrd}
\end{figure}

Finally, the effects of outshining may shed some light on a long-standing tension between observations of the stellar mass density (SMD) of the universe and the star-formation rate density (SFRD). The SMD at any time should be the integral of the SFRD corrected for mass lost and returned to the interstellar medium. A large body of work has reported a discrepancy between the inferred SFRD and the directly measured one, particularly at higher redshifts. \citet{Wilkins2008} found instantaneous indicators of the SFRD were 0.6 dex higher than those fit to the stellar mass history at $z > 2$. \citet{Ilbert2013} compared their inferred star formation history to data compiled by \citet{Behroozi2013} and found a difference of 0.2 dex at $z > 1.5$ (although the discrepancy was still within the expected uncertainties). \citet{Huertas-Company2016} recently inferred a star formation history extremely similar to \citet{Ilbert2013} despite a completely different dataset, but, when they compared their results to a recent compilation of direct SFRD measurements by \citet{Madau2014}, they found the direct measurements were $\sim$ 1.25 times larger than their inferred SFRD at $z > 2$ ({\color{black}blue stars} versus broken black line in Figure \ref{fig:sfrd}). \citet{Reddy2009} and \citet{Sawicki2012a} attributed this difference to poor constraints on the low mass tail of the stellar mass functions, but the issue has persisted even with more recent and deeper surveys, and an alternate explanation may be required.  

To test how outshining would affect SMD measurements, we first took the best-fit Schecter function parameters for all galaxy morphologies listed in Table 2 of \citet{Huertas-Company2016}, which were divided at different redshifts into star-forming and passive mass functions. We estimated the star-formation rate of the star-forming galaxies {\color{black} of a given mass and redshift} using the main sequence relation of \citet{Johnston2015}.
%, whose parameterization allowed us to calculate SFRs easily at any redshift. 
Then, {\color{black} with estimates of} the sSFRs of the star-forming population of galaxies, we {\color{black} used Equation \ref{eq:massCorrection} to correct} the star-forming mass function at each redshift for the mass bias found in this work. Following \citet{Huertas-Company2016}, we integrated below the {\color{black}corrected} curve at each redshift between 10$^8$ and 10$^{12}$ \Msun, and added the integral of the passive mass function {\color{black} in that same mass range} to find the SMD as a function of redshift. Following the procedure laid out by \citet{Wilkins2008}, we related the {\color{black} corrected} SMD to the SFRD using
\begin{equation}\label{eq:smd}
  SMD = \int_0^tSFRD(t')(1 - 0.05\ln(1 + \frac{t - t'}{0.3 \mathrm{Myr}}))dt'
\end{equation}
where the parameterization for the return fraction is assumed to be that given by \citet{Conroy2008} for a Chabrier IMF.

The functional form of the SFRD is taken to be that given by \citet{Behroozi2013}, namely
\begin{equation}
  SFRD(z) = \frac{C}{10^{A(z - z_0)} + 10^{B(z - z_0)}}.
\end{equation}
The SFRD was fit holding $A$ fixed at $-1$ as done in the previous works. {\color{black} The parameters $B$, $C$, and $z_0$ were determined through $\chi^2$ minimization of Equation \ref{eq:smd}, yielding a redshift-dependent SFRD inferred from observations of stellar mass functions.} Our best fit parameters were $B = 0.15$, $C = 0.12$, and $z_0 = 0.99$ {\color{black} using the $z_s + z_p$ correction}. Because it is not our intent to actually measure the star formation history of the universe, only show how a pixel-by-pixel mass correction would affect previous results, we have refrained from performing any uncertainty analysis on these parameters.

The results of {\color{black} inferring the SFRD of the universe} are shown in Figure \ref{fig:sfrd}. The dashed black curve shows the original best-fit star formation history of \citet{Huertas-Company2016}, and the most recent compilation of direct SFRD measurements from \citet{Madau2014} is shown as blue stars. The discrepancy between these two different measurements is evident, particularly above $z \sim 1.5$. Our correction to the inferred star formation rate history is shown as the yellow and green curves for fits to the $z_s$ and ($z_s + z_p$) XDF samples respectively. They follow the original curve of \citet{Huertas-Company2016} closely at low redshift, but begin to diverge significantly around $z > 1.5$, and become more inline with the direct SFRD measurements. There is still a maximum offset of 0.03 dex at $z \sim 2$ between our $z_s + z_p$ correction and the direct measurements, and the slope at the highest redshifts may be too shallow compared to that found by \citet{Madau2014}. These differences may arise due to the different functional parameterization used by \citet{Madau2014}, or due to our assumed conversion between a galaxy's mass and SFR given by \citet{Johnston2015}, which had a higher sSFR for lower mass galaxies than the relation from \citet{Whitaker2014}. In general, it is clear that the mass correction presented in this work acts to increase the SMD at higher redshifts, and thus greatly reduce the tension between inferred (from stellar mass functions) star formation histories and directly measured ones. 

{\color{black}We perform a complementary test } to see how close a mass correction of the same piecewise form {\color{black} given by Equation \ref{eq:piecewiseFit}} could come to the direct observations. {\color{black} For this,} we converted the \citet{Madau2014} SFRD into a SMD {\color{black} as a function of redshift. We then} found the best piecewise parameters that would make the \citet{Huertas-Company2016} mass functions match this SMD. {\color{black} Note that this is in a sense the reverse process of that given earlier in this section: instead of having corrected stellar mass functions and fitting a functional form of the SFRD, we now are given a functional form of the SFRD from \citet{Madau2014} and are going to determine what correction needs to be applied to the stellar mass functions.} The resulting best correction is shown in Figures \ref{fig:hizmasscompwithnoise} and \ref{fig:sfrd} as the solid red curve, and it is generated using the parameters ${p, m, b, k_2} = {-10.1, -0.02, -0.03, -0.06}$. The best-fit correction to match the directly measured SFRD has a linear component that is within uncertainty of the XDF fits, but a break-point ($p$) that lies just to the left the 1-$\sigma$ confidence region, and a parabolic component that is shallower than would be expected from the XDF data. The general similarity between the form of the best-fitting correction and the XDF corrections (if not the exact details) lends further credence to the observed kink in the mass discrepancy at high sSFRs. Overall, we find that correcting the observed stellar mass functions for mass missed by unresolved SED fitting eliminates the ``missing mass'' problem that is otherwise seen when comparing direct measurements of the SFRD with those inferred from the SMD.

A potential caveat regarding these results is that the Hubble XDF is a small slice of the sky compared to the scale that most stellar mass functions are measured from (4 $arcmin^2$ compared to 880 $arcmin^2$ for \citet{Huertas-Company2016}, or 1.62 $deg^2$ for \citet{Muzzin2013}). The (relatively) small number of galaxies limit how sweeping our conclusions can be. For example, there are only a small number of galaxies (55) with masses between $10^{10.4}$ and $10^{11.4} \Msun$, which corresponds roughly to plus or minus three times $M^*$. While the $M^*$ galaxies in our sample all follow the trend shown in Figure \ref{fig:hizmasscompwithnoise}, they all have sSFRs that lie to the left of the break point, meaning it is unclear if the population of $M^*$ galaxies would also turn sharply downward. Fortunately, the XDF does include a large number of sub-$M^*$ galaxies, which is where the bulk of stellar mass lies (approximately 80\% as shown in Figure 1 of \citet{Sawicki2006b} for a faint end slope $\alpha$ of -1.4, consistent with deep SMF estimates from \citet{Tomczak2014}). In order to study how pixel-by-pixel mass estimates perform with varying galaxy populations, greater numbers are needed, and must wait for the onset of next generation high-resolution wide-field astronomical imagers, such as WFIRST.

\subsection{Why was this bias not seen before?}
\label{sec:why}

As discussed above, \citet{Sorba2015} observed a bias in stellar mass measurement due to outshining for nearby galaxies at low redshift. This work shows a similar trend for galaxies up to $z = 1.2$, and a more egregious offset in mass for galaxies from $1.2 < z < 2.5$. However, previous pixel-by-pixel analysis at high redshift by \citet{Wuyts2012} found zero mass discrepancy when comparing the pixel-by-pixel and unresolved mass estimates for star-forming galaxies galaxies at $0.5 < z < 1.5$ and $1.5 < z < 2.5$. Although their selection criteria differed from ours (they only performed pixel-by-pixel SED fitting on galaxies with masses greater than $10^{10}$ \Msun\ and log sSFRs greater than -9.76 yr$^{-1}$ and -9.51 yr$^{-1}$ for their low-$z$ and high-$z$ samples respectively), our results indicate that they should have seen an unresolved versus resolved mass bias regardless. It is necessary to understand what led to our disparate results. Perhaps the most obvious factor that could have contributed to the outshining bias not being measured is that it is a small effect, at least for galaxies with lower sSFRs. When averaged over all sSFRs, our XDF low redshift ($z \leq 1.2$) spectroscopic sample displays a mean bias of only 0.04 dex. It is only the increasing difference with sSFR that makes the bias particularly notable. Nevertheless, here we examine the differences in our methodologies to see what other factors would influence the unresolved to pixel-by-pixel mass ratios.

The first difference is \citet{Wuyts2012} constructed their model SED templates with \citet{BC03} models, whereas we used FSPS models. Although the manner in which various codes treat certain phases of stellar evolution such as the Thermally Pulsating Asymptotic Giant Branch (TP-AGB) can have strong effects on \mtl ratios, this should not affect the mass ratios in each work, as both the pixels and integrated light are fit consistently within that work.

Second, \citet{Wuyts2012} used a modified $\chi^2$ statistic which incorporated additional integrated fluxes ($U$, $K_s$, and the four IRAC bandpasses) in order to find the best fitting model. The statistic they minimized was
\begin{equation}
  \chi^2_{res+int} = \chi^2_{res} + \sum_{j=1}^{N_{int}} \frac{(F_j - \sum_{i=1}^{N_{bin}}M_{i,j})^2}{E_j^2}
\end{equation}
where $\chi^2_{res}$ is the sum of all the $\chi^2$ values as normally found in SED fitting for each spatial bin (essentially each pixel) using only bandpasses that are fully spatially resolved, $N_{int}$ is the number of spatially-integrated-only bandpasses, $N_{bin}$ is the number of spatial bins, $F_j$ and $E_j$ are the observed flux and error in the $j^{th}$ integrated only bandpass, and $M_{i,j}$ the model flux in the $j^{th}$ integrated only bandpass and $i^{th}$ spatial bin. Because the parameter space of this statistic is very large, they started with an initial guess using the best-fit model of each spatial bin found using only the resolved bandpasses and then iteratively adjusted one spatial bin's model at a time to improve the $\chi^2_{res+int}$. This iteration was continued until no improvement was found or a maximum of 500 iterations. In contrast, our pixels were treated essentially as their own, independent, objects. This would seem to be a major difference in our approaches, but \citet{Wuyts2012} clearly state that the good correspondence between pixel-by-pixel and unresolved mass estimates remains even if constraints from integrated fluxes were ignored.

\begin{figure}
  \includegraphics[width=\columnwidth]{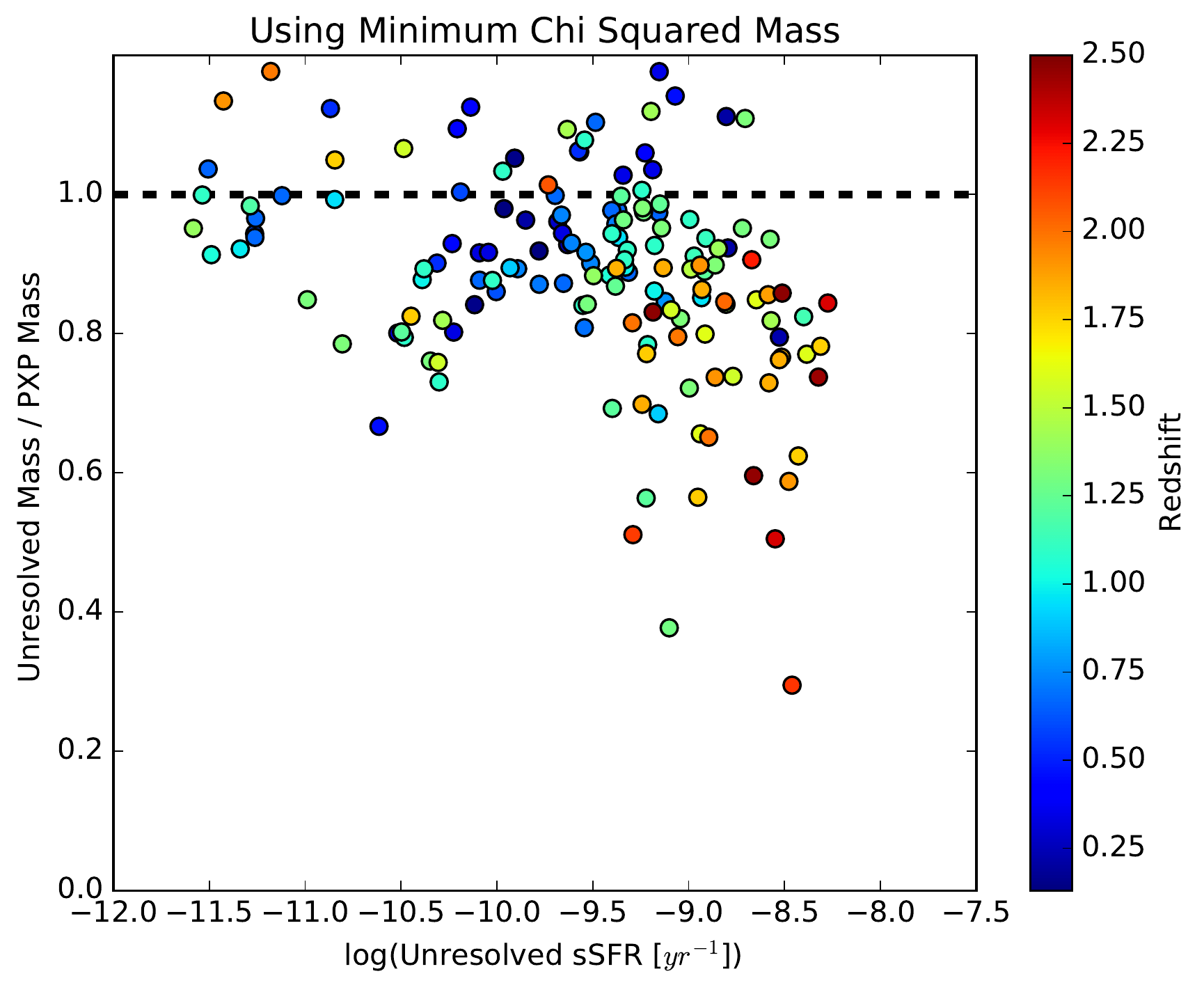}
  \caption{{\color{black} Same as the left panel of Figure \ref{fig:masscomp}, but showing results of using the minimum $\chi^2$ parameter estimates. The linear trend at lower sSFRs is de-emphasized and the mass discrepancy plummets more abruptly with several galaxies still remaining near the one-to-one line at higher sSFRs.}}
  \label{fig:chi2masscomp}
\end{figure}

Perhaps their choice to use the model with the minimum $\chi^2$ as their best-fit mass estimate versus our use of the median mass from several hundred Monte Carlo iterations makes a difference. \citet{Taylor2011} have shown how using the best-fitting model to define the stellar mass can induce strong systematic effects. A simple check of finding the mass ratios of our best-fit model masses (rather than median masses) for the spectroscopic sample gives an average mass ratio of $0.94 \pm 0.10$ which is higher than the average of the median mass ratios of $0.91 \pm 0.06$, but not significantly. {\color{black} More telling, as shown in Figure \ref{fig:chi2masscomp} the minimum $\chi^2$ mass ratios show a lack of a linear trend with sSFR, with a slope of only $-0.014 \pm 0.11$, until sharply dropping at practically a right angle. The downward plunge in mass ratio is much more sudden, and driven by the high sSFR galaxies, while the rest remain close to the one-to-one line. This means that for a random selection of galaxies, the increasing bias with sSFR would not be as obvious when using mass derived from minimized $\chi^2$ fits, and emphasizes the value of using the median mass of many perturbed SED fits. However, by selecting galaxies with only high sSFRS, the overall bias should still be present in the sample used in \citet{Wuyts2012}, even if the minimum $\chi^2$  masses would not show the bias as strongly as median mass estimates. } 

Finally, in order to increase signal-to-noise in their spatial bins, \citet{Wuyts2012} combine pixels together using the Voronoi two dimensional binning technique of \citet{Cappellari2003}. The pixel binning degrades their spatial resolution, which is a key factor in allowing pixel-by-pixel analysis to uncover any bias due to outshining. The pixel-by-pixel SED fitting allows one to separately fit spatially segregated stellar populations, thus better constraining the \mtl ratio of the galaxy built up over its entire star-forming history. As the spatial resolution gets poorer and poorer, the ability to see fainter stars behind the bright stars necessarily goes away. {\color{black} \citet{Sorba2015} studied the effects of resolution on pixel-by-pixel mass estimates and found that a minimum spatial resolution of 2-3 kpc was required to reveal a mass discrepancy. At the pixel scale of the XDF (60 mas) and redshift two, this would mean binning together 16 to 25 pixels would be enough obscure the effects of outshining, an amount commonly needed to meet the signal-to-noise criteria of 10 required by \citet{Wuyts2012}, particularly in the intermediate and outer regions of galaxies.  } It is likely then, that the spatial binning performed by \citet{Wuyts2012} pushed the pixel-by-pixel mass estimates into a regime where outshining could not be measured. In fact, when they compared resolved to unresolved mass estimates without binning, they found the pixel-by-pixel masses were systematically heavier by 0.2 dex (about 27\%). They state that the binning to a minimum signal-to-noise is necessary to avoid biases in SED fits that lead to erroneously large \mtl ratios. This is a true concern as discussed above, exacerbated by their use of minimum $\chi^2$ best-fits. Their use of binning was entirely justified, especially considering the focus of their analysis on star-forming clumps and overall surface profiles as a function of half-mass/half-light radii. The binning did, however, along with the use of minimum $\chi^2$ masses, act to obscure evidence of the outshining bias found in our work.

\section{Conclusions}
\label{pxp2:conclusions}

{\color{black} 
In this work we fit spatially-resolved galaxies in the Hubble XDF with spectral synthesis models. We focused on comparing galaxy stellar mass estimates from spatially-resolved model fits to those from spatially-unresolved, integrated light. Our key result is that spatially-unresolved fitting underestimates galaxy masses compared to the more realistic, spatially-resolved fits.  For high-sSFR galaxies the discrepancy can be large, as large as a factor of $\sim$5 in some cases. 

In detail, our findings are: 
}

\begin{itemize}
\item Following \citet{Sorba2015}, we examined the ratio of the unresolved mass estimate to the pixel-by-pixel mass estimate as a function of sSFR. We found that a bias in the mass estimate of galaxies due to outshining still persists up to $z = 2.5$. Moreover, the bias appears to have two distinct components:
  \begin{enumerate}
  \item At low sSFRs (less than approximately $10^{-9.2} yr^{-1}$), the ratio of unresolved to pixel-by-pixel mass changes slowly and roughly linearly with log(sSFR). The slope and intercept of a linear fit is commensurate with the results found for nearby galaxies in \citet{Sorba2015}.
  \item Above sSFR $= 10^{-9.2} yr^{-1}$, the mass ratio turns sharply downward, and can no longer be represented by the shallow linear trend found at lower redshifts. Because sSFRs tend to be larger at higher redshift, the sharply increasing bias implies that masses for some galaxies between redshifts 1.5 and 2.5 could systematically be underestimated by factors of 2--5.
  \end{enumerate}
  
  \item The piecewise form of the mass correction (shallow at low sSFRs and becoming much steeper as sSFR increases) naturally resolves a long-standing tension between the directly observed SFRD and that derived from the observed SMD. Correcting stellar mass functions from \citet{Huertas-Company2016} brings the derived SFRD much more in line with the directly observed SFRD compilation of \citet{Madau2014}.
  \item Correcting mass estimates for the bias found above slightly alters the slope of the star-forming main-sequence. It also reduces the intrinsic scatter about the star-forming main-sequence, suggesting that star-formation in galaxies proceeds in a smooth fashion.
  
\end{itemize}

{\color{black}
Spatially resolved SED fitting is a powerful tool, and differences between resolved and unresolved fits highlight the importance of treating galaxies as composite structures, rather than one cohesive whole.

In practical terms, the more realistic, spatially-resolved fits can give significantly higher masses than the unresolved fits that are necessarily used with unresolved, ground-based data.  We recommend that studies using unresolved fits correct their stellar mass estimates using a formula such as the one we provided in Eq.~\ref{eq:massCorrection}. 
}

\section*{Acknowledgements}

We thank the anonymous referee for many insightful comments that greatly improved the quality of this paper.
We are indebted to the high-quality data products made publicly available by the Hubble XDF Team. We are very grateful to Anneya Golob, Liz Arcila-Osejo, and Adam Muzzin for useful discussions related to this work. RS wishes to thank his colleagues at both institutions for their support.

This work was supported financially by the Natural Sciences and Engineering Research Council (NSERC) of Canada, including an NSERC Graduate Scholarship to RS and an NSERC Discovery Grant to MS. 

Computational facilities are provided by ACEnet, the regional high performance computing consortium for universities in Atlantic Canada. ACEnet is funded by the Canada Foundation for Innovation (CFI), the Atlantic Canada Opportunities Agency (ACOA), and the provinces of Newfoundland and Labrador, Nova Scotia, and New Brunswick. 

This research made use of Astropy, a community-developed core Python package for Astronomy \citep{Robitaille2013}, as well as many of the routines available in the scipy and numpy software packages.

%%%%%%%%%%%%%%%%%%%%%%%%%%%%%%%%%%%%%%%%%%%%%%%%%%



\bibliographystyle{mnras}
\bibliography{pxpxdf} % if your bibtex file is called example.bib

%\begin{thebibliography}{99}
%\end{thebibliography}

%%%%%%%%%%%%%%%%%%%%%%%%%%%%%%%%%%%%%%%%%%%%%%%%%%

%%%%%%%%%%%%%%%%% APPENDICES %%%%%%%%%%%%%%%%%%%%%

\appendix

\section{Model Parameter Space Study}

\begin{figure*}
  \includegraphics[width=\textwidth]{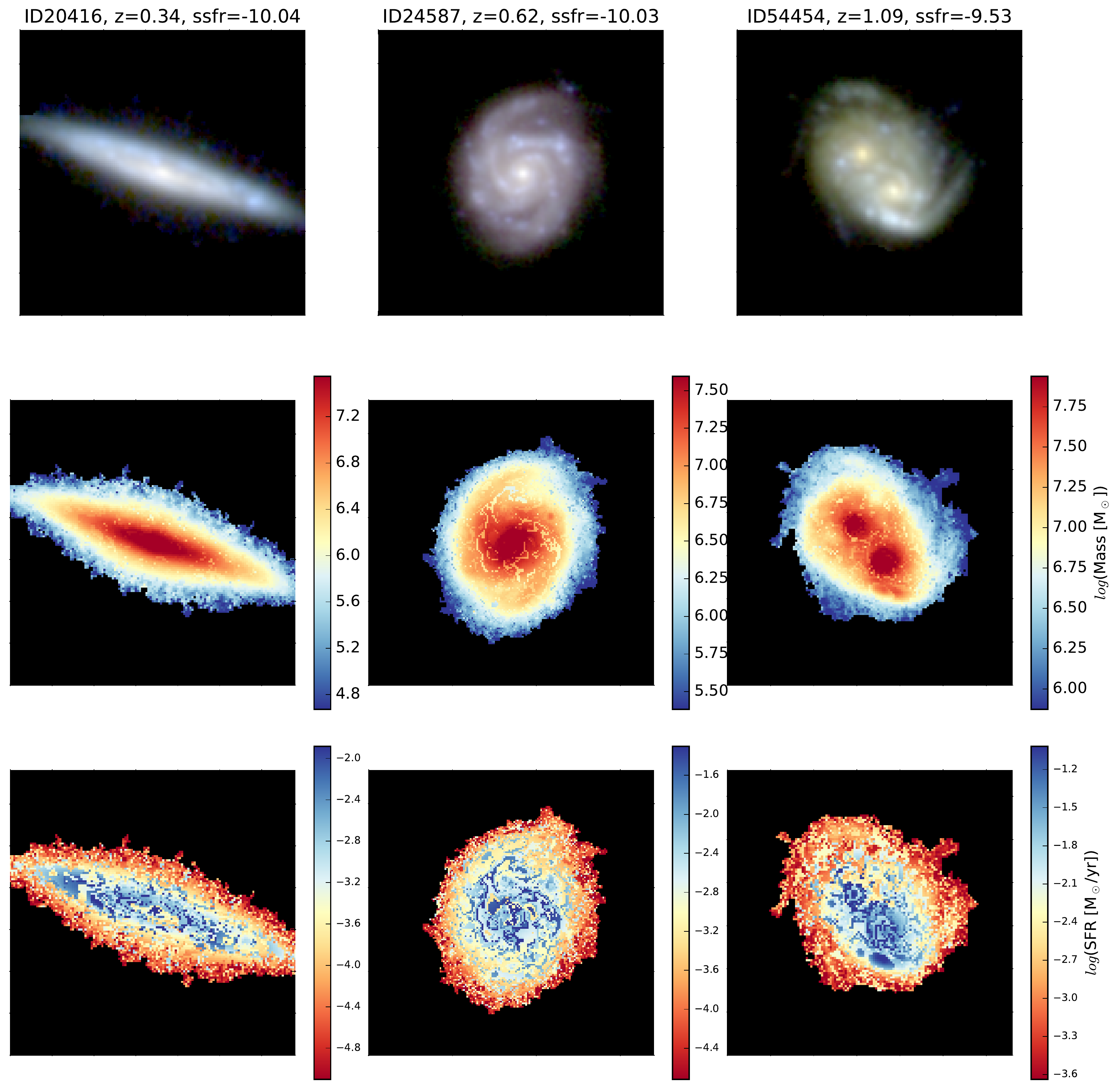}
  \caption{{Two-dimensional maps of three representative XDF galaxies created using the same model SED set as \citet{Sorba2015}. From top to bottom we show false-color images in approximately rest-frame $ugr$, stellar mass, and SFR. This model set did not provide adequate pixel-by-pixel SED fitting. The galaxies have, from left to right, $RA = \{53.1712061^{\circ}, 53.1699419^{\circ}, 53.14785833^{\circ}\}, Dec = \{-27.81471020^{\circ}, -27.7710194^{\circ}, -27.77403611^{\circ}\}, z_{spec} = \{0.337, 0.622, 1.088\},$ and $\mathrm{log}{(M_*/\Msun)} = \{10.3, 10.7, 10.9\}$ where the mass measurement is the median pixel-by-pixel mass from our final catalog.}}
  \label{fig:burstmap}
\end{figure*}

{\color{black} To state the obvious, the parameter estimates derived from SED fitting depend on the model template grid used; a poorly suited model grid will give poor results. This is true with pixel-by-pixel SED fitting as well, but because  the pixel-by-pixel allows for the creation of 2D property maps, we have an extra check on the validity of our models that is not possible when fitting the entire galaxy's light as one.} This is shown in Figure \ref{fig:burstmap} where we examine property maps created when using two-component burst SFHs as in \citet{Sorba2015}. We can see the two-dimensional stellar mass and SFR distributions are not at all smooth as would be expected from the false-colour images. The stellar mass maps contain wispy filaments of pixels much less massive than the surrounding stellar material, and the SFR maps show no discernible structure when we would expect them to roughly trace the blue stars in the spiral arms. Why two-component burst models would work well for SDSS galaxies, but fail for XDF galaxies is an open question. It could be that the SDSS observations probe smaller physical scales, but it may also be related to the results of \citet{Gallazzi2009}, who found that including a large fraction of bursty models could under-estimate the mass-to-light ratio of galaxies with smoother SFHs. As we observe galaxies at greater redshifts, it becomes more likely that they will have had smoother SFHs, as any recent bursts will have stellar populations not too dissimilar from the underlying older stellar populations.

It could also be that the two-component burst models are too degenerate at higher redshifts. Specifically, the mass fraction of the young stellar component can not be constrained by the observations, and not only could the models fail to capture a true underlying older population of stars, they could erroneously add a large fraction of mass to a strictly younger population of stars by over-estimating the older stellar component. This would not be the case for low redshift galaxies in the SDSS, since it is very unlikely to observe a pixel actively forming stars that did not have an underlying older stellar component. The models may over-fit the older component somewhat at low redshift, but not nearly as drastically as is possible at higher redshifts, where there has not been as much time to assemble mass in the older population. The over-estimation of the older stellar component would also affect SFR measurements since each component's SFH is modelled as an exponential decay. This would explain the poor reproduction of the star-forming regions seen in the bottom row of Figure \ref{fig:burstmap}. To ensure realistic spatially resolved stellar property maps, we use one-component $\tau$-models, which were shown to give qualitatively similar mass estimates as two-component burst models for galaxies in the SDSS \citep{Sorba2015}. Note that with one-component models, outshining will always act to hide the mass of older stars.

\begin{figure*}
  \includegraphics[width=\textwidth]{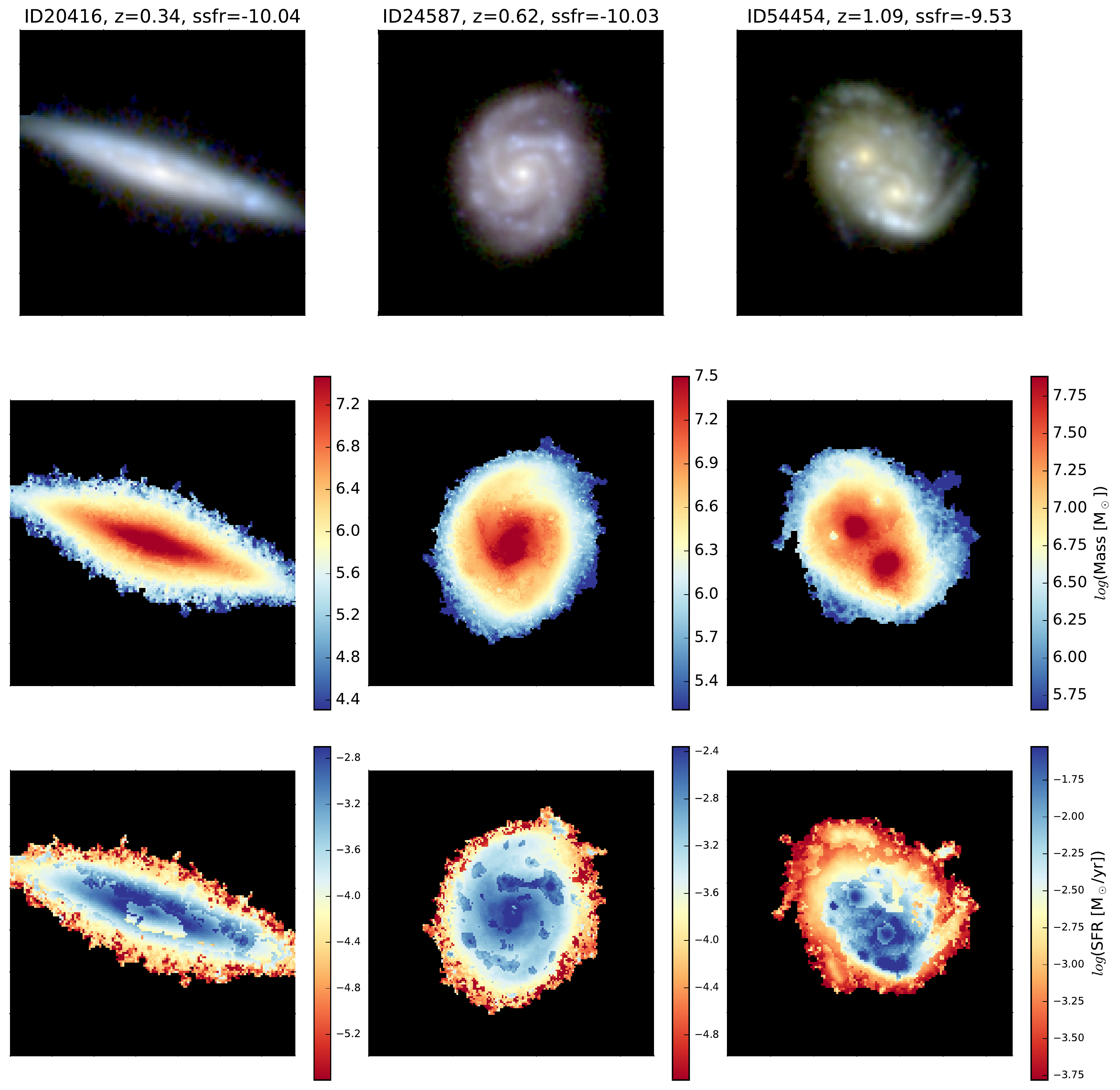}
  \caption{{Same as Figure \ref{fig:burstmap}, except created using a coarse grid of $\tau$-models. These show a slight improvement over the burst models.}}
  \label{fig:ctaumap}
\end{figure*}

We found that it is imperative to have a parameter space that is well sampled with regards to SFH when doing pixel-by-pixel SED fitting. Figures \ref{fig:ctaumap} and \ref{fig:ftaumap} show stellar mass and SFR maps when using coarse and fine model grids respectively. The coarse grid contains six different values for $\tau$, ranging from 0.3 to 10 Gyr and the spacing between each step roughly doubling each time, and ten different ages between $10^{7.75}$ and $10^{10}$ years, with the exponent increasing in steps of 0.25. As can be seen in Figure \ref{fig:ctaumap}, although the stellar population property maps are smoother than those of the burst models, the SFR map in particular still shows sharp discontinuities that seem unrelated to any features visible in the false-color images. These discontinuities disappear when using finer spacings for the SFH parameters, as shown in Figure \ref{fig:ftaumap} of the main text. Here the model grid consists of 20 different $\tau$ values spaced roughly evenly logarithmically between 0.3 and 10, and log(ages/yr) ranging from 7.7 to 10.1 in steps of 0.05. The finer model grid displays much smoother property maps with no unphysical discontinuities. It is this set of models that we chose to use as our ultimate template grid. Note that the need for a finely spaced model grid may also have played a role in the poor performance of the two-component models above. However, because two-component models have many more free parameters, the total number of models grows quickly as the resolution of each parameter is increased, and can become cumbersome to work with.

Studying the right-most mass map of Figure \ref{fig:ftaumap} we can see an example of outshining happening on a small scale as evidenced by the two ``dimples'' of less massive pixels to the left and right of the upper bulge. It is unlikely that there are massive holes in the galaxy when the rest of the mass map shows a very smooth gradient. These two areas correspond to the bright blue star-forming regions visible in the false-color image, indicating that the youngest stars are obscuring an older stellar population, leading to an erroneous \mtl ratio. Similar dimples can also be seen for the brightest blue spots in the spiral arms of the left-most galaxy. This effect seems unavoidable with $\tau$ models, and even in SED fitting general since the light from very different SFHs can be dominated by the younger stellar populations. With no easy way to distinguish between SFHs with different \mtl ratios, noisy observations can easily be fit to models that underestimate the stellar mass, even if the correct SFH was present in the model grid. We do not correct our pixel-by-pixel masses for this dimpling effect, but estimating the missing mass by interpolating the surrounding pixels shows that any correction would be of order less than 1\% of the total mass of the galaxy.

\begin{figure*}
  \includegraphics[width=\textwidth]{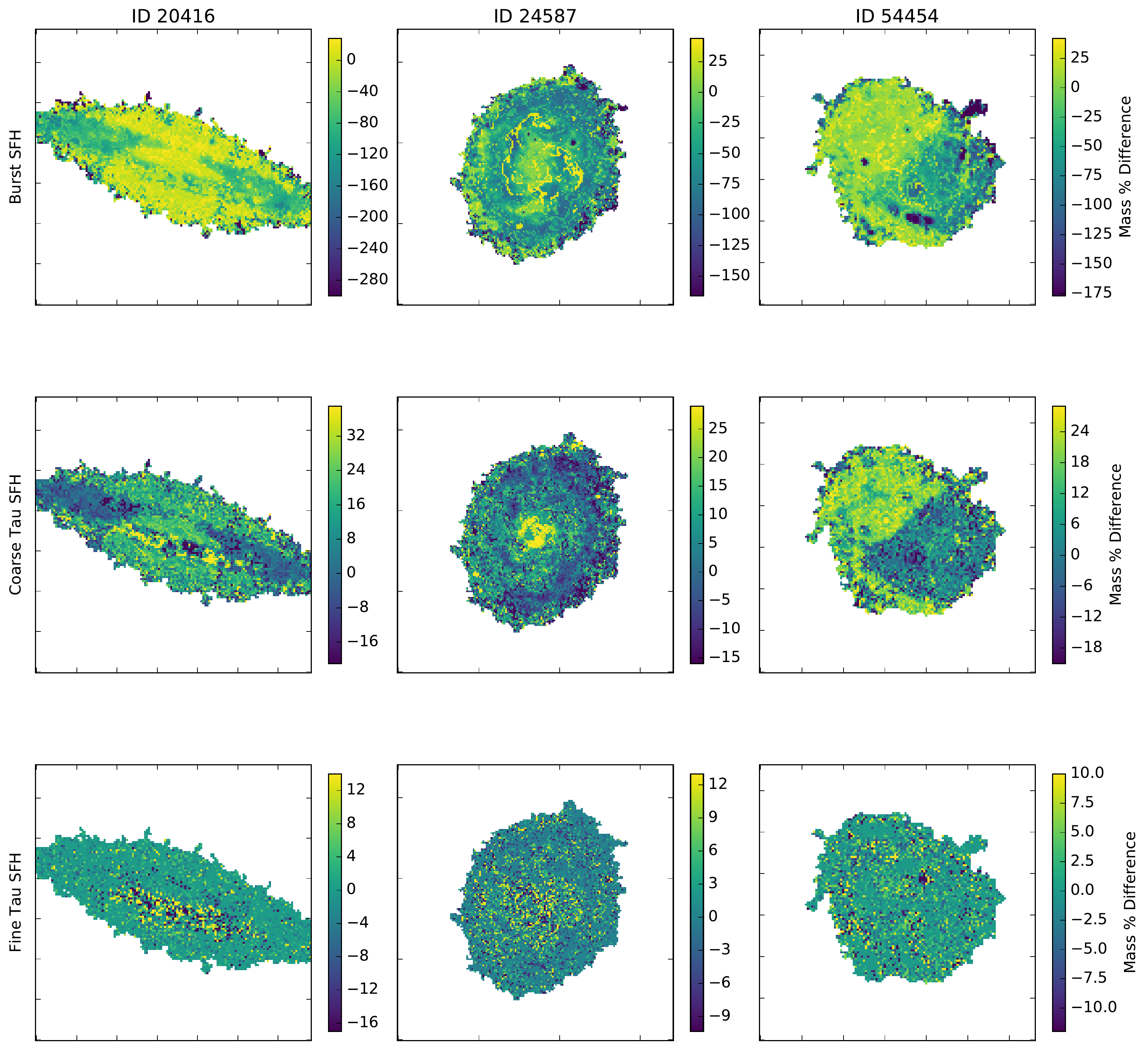}
  \caption{{Relative residual mass maps created by dividing the mass map of the specified model grid (burst SFH in the top row, coarse tau SFH in the middle, and fine resolution tau in the bottom) by the ``best'' mass map created with the doubly-fine tau SFH grid. The results for three galaxies are shown as columns, demonstrating that the fine grid has roughly the same results as the doubly-fine.}}
  \label{fig:doublyfine}
\end{figure*}

{\color{black} To ensure that our model grid had adequate resolution, we created an even finer set of model templates by doubling the number of values used for each parameter (tau, age, and dust) while keeping the upper and lower limits the same. We compared the mass maps made with this doubly-fine template grid to those of the other template grids by taking the ratio. These residual maps are shown in Figure \ref{fig:doublyfine} for our three test galaxies. As expected, the burst (top row) and coarse model (middle row) grids show strong spatial correlations in the residuals as well as large relative differences. The fine model grid (bottom row), however, compares well to the doubly-fine grid, with the residuals looking more like random noise and relative differences typically less than approximately 10\%. As a further check that the quality of fits had converged, we investigated how the average $\chi^2$ value of the best-fitting model changed with our model grid. These values are shown in Table \ref{tab:doublyfine}. Of particular note is that the fine and doubly-fine values are virtually equivalent, demonstrating that our chosen resolution is enough to represent the galaxy pixels.}

\begin{center}
\begin{table}

\centering
\caption{\label{tab:doublyfine} Average $\chi^2$ values of various model grid templates fit to pixels of three different galaxies.}
\begin{tabular}{@{}lcccc}
\hline
Galaxy ID &  & $\chi^2$ &  \\
 & Burst & Coarse $\tau$ & Fine $\tau$ & Doubly-Fine $\tau$ \\ 
\hline
20416 & 17.59 & 13.80 & 13.70 & 13.60 \\
24587 & 23.99 & 9.24 & 8.07 & 8.06 \\
54454 & 16.37 & 13.27 & 11.36 & 11.21 \\
\hline
\end{tabular}
%\end{centering}
\end{table}
\end{center}

%%%%%%%%%%%%%%%%%%%%%%%%%%%%%%%%%%%%%%%%%%%%%%%%%%

% Don't change these lines
\bsp	% typesetting comment
\label{lastpage}
\end{document}